\documentclass[%
 reprint,
superscriptaddress,
 amsmath,amssymb,
 aps,
]{revtex4-2}

\usepackage{graphicx}
\usepackage{dcolumn}
\usepackage{bm}

\usepackage{braket}
\usepackage{soul}  

\usepackage[hidelinks]{hyperref}
\usepackage{graphicx}
\usepackage{capt-of}
\usepackage{xcolor}
\hypersetup{
  colorlinks   = true, 
  urlcolor     = blue, 
  linkcolor    = blue, 
  citecolor   = blue 
}

\def\E_r{E_{\text{r}}}
\def\k_r{k_{\text{r}}}
\def\t_hold{t_{\text{hold}}}
\def\omegaR{\omega_{\text{R}}}
\def\deltaR{\delta_{\text{R}}}

\begin{document}

\preprint{APS/123-QED}

\title{Bose-Einstein Condensate on a Synthetic Topological Hall Cylinder}

\author{Chuan-Hsun Li}
\affiliation{School of Electrical and Computer Engineering, Purdue University, West Lafayette, Indiana 47907, USA}
\affiliation{Department of Physics and Astronomy, Purdue University, West Lafayette, Indiana 47907, USA}
\author{Yangqian Yan}
\email{Current affiliation: Department of Physics, The Chinese University of Hong Kong, Shatin, New Territories, Hong Kong, China}
\affiliation{Department of Physics and Astronomy, Purdue University, West Lafayette, Indiana 47907, USA}
\author{Shih-Wen Feng}
\author{Sayan Choudhury}
\author{David B. Blasing}
\affiliation{Department of Physics and Astronomy, Purdue University, West Lafayette, Indiana 47907, USA}
\author{Qi Zhou}
\email{Corresponding author. zhou753@purdue.edu}
\affiliation{Department of Physics and Astronomy, Purdue University, West Lafayette, Indiana 47907, USA}
\affiliation{Purdue Quantum Science and Engineering Institute, Purdue University, West Lafayette, Indiana 47907, USA}
\author{Yong P. Chen}
\email{Corresponding author. yongchen@purdue.edu}
\affiliation{Department of Physics and Astronomy, Purdue University, West Lafayette, Indiana 47907, USA}
\affiliation{School of Electrical and Computer Engineering, Purdue University, West Lafayette, Indiana 47907, USA}
\affiliation{Purdue Quantum Science and Engineering Institute, Purdue University, West Lafayette, Indiana 47907, USA}

%
\begin{abstract}
The interplay between matter particles and gauge fields in physical spaces with nontrivial geometries can lead to novel topological quantum matter. 
However, detailed microscopic mechanisms are often obscure, and unconventional spaces are generally challenging to construct in solids.
Highly controllable atomic systems can quantum simulate such physics, even those inaccessible in other platforms. Here, we realize a Bose-Einstein condensate (BEC) on a synthetic cylindrical surface subject to a net radial synthetic magnetic flux. We observe a symmetry-protected topological band structure emerging on this Hall cylinder but disappearing in the planar counterpart. BEC's transport observed as Bloch oscillations in the band structure is analogous to traveling on a M\"obius strip in the momentum space, revealing topological band crossings protected by a nonsymmorphic symmetry. We demonstrate that breaking this symmetry induces a topological transition manifested as gap opening at band crossings, and further manipulate the band structure and BEC's transport by controlling the axial synthetic magnetic flux. Our work opens the door for using atomic quantum simulators to explore intriguing topological phenomena intrinsic in unconventional spaces.

\end{abstract}

\maketitle


\section{\label{sec:introduction} Introduction}

Physical spaces with nontrivial geometries can give rise to novel phenomena difficult to attain in planar spaces.
Such unconventional spaces are studied in various disciplines such as general relativity and cosmology \cite{Spacetime_geometry_carroll_2019}, photonics \cite{Synthetic_Landau_level_Nature2016,Curvedspace_photonic_lattice_PhysRevA2017,control_light_curved_Bekenstein_Natphys2017, gravitational_responses_photonic_Landau_Nature2019,Hyperbolic_lattice_Nature_2019}, 
and condensed matter physics \cite{Gauge_Fields_Condensed_Matter_1989,Vortices_curvedspace_RevModPhys2010,Dirac_eqn_curved_Boada_2011,Gauge_field_curvedspace_Zhai_JPB2017}.
For example, gravity stems from curved spacetimes in general relativity \cite{Spacetime_geometry_carroll_2019}.
Superfluids on curved surfaces 
carry vortices with no counterpart in planar spaces \cite{Cylinder_Ho_PhysRevLett2015,Vortex_cylindrical_PRA2017}.
Fractional quantum Hall states become degenerate on a torus, underlying the profound concept of topological order \cite{Degeneracy_FQH_PhysRevB1990}. 
Highly controllable atomic systems offer opportunities to quantum simulate
\cite{Quantum_simulation_RevModPhys2014,Quantum_Simulators_Architectures_PRXQuantum2021} various phenomena and uncover new physics inherent in unconventional spaces, 
including those challenging to study in conventional platforms.
For instance, 
various analogues of cosmic phenomena \cite{Hawking_entanglement_Steinhauer_2016,Expanding_ring_BEC_PhysRevX2018,Unruh_radiation_Hu2019}
have been observed in table-top experiments with a Bose-Einstein condensate (BEC),
in which excitations such as phonons are in an effective curved spacetime
\cite{probe_gravity_bec_PhysRevA2003,Gibbons_Hawking_PhysRevLett2003}.
Superfluid properties can be studied in detail with a BEC prepared in 
a ring trap \cite{Quantized_supercurrent_ring_BEC_PhysRevA2012,Hysteresis_superfluid_ring_Nat2014}.

Topological quantum matter \cite{Hasan_TI_RevModPhys,Qi_TI_TS_RevModPhys,Classification_topological_symmetry_RMP2016,Zoo_topological_matter_Wen_RevModPhys2017} 
has received great attention across different areas 
because of its robust properties promising for reliable devices and quantum information processing
\cite{Topological_quantum_computation_review_science2013}.
Whereas topological phenomena in planar spaces have been studied extensively,
those intrinsic in unconventional spaces remain substantially unexplored in experiments,
because it is challenging to realize spaces that simultaneously incorporate the underlying novel geometries with crucial ingredients such as gauge fields 
and to manipulate the required Hamiltonians.
For instance, 
threading a magnetic flux through a two-dimensional (2D) plane 
has led to the remarkable discovery of the quantum Hall effects and various topological quantum matter 
\cite{Hasan_TI_RevModPhys,Qi_TI_TS_RevModPhys,Classification_topological_symmetry_RMP2016,Zoo_topological_matter_Wen_RevModPhys2017} for electrons. 
However, creating a net radial magnetic flux through the cylindrical surface of a nanotube is extremely difficult. 
Such a Hall cylinder is an important paradigm for many theoretical studies of topological physics \cite{Laughlin_pump_PhysRevB1981,Interaction_PBC_NatCom_2015,Fractional_QH_real_cylinder_PRA_2016, Fractional_pump_synthetic_cylinder_PRL_2017,Yan_PRL_2018,Zhang_Hall_Tube_PhysRevA2020}, but its experimental exploration is largely lacking.
The microscopic mechanisms of how changing the geometry of the underlying space may give rise to distinct topological matter require further research.
 
Atomic quantum simulators, such as atoms in optical lattices \cite{Bloch_RevModPhys2008,Bloch_Nature_review_2012,QuantumSimulationLattice_review_science2017,QuantumSimulationLattice_review_science2017,Tools_quantum_simulation_OL_NatRevPhys_2020}
subject to additional ingredients like synthetic gauge fields 
\cite{Dalibard_gaugefield_RevModPhys,Spielman_Review,ZhangReview2014,Zhai_Review,YJLin_Review,Engels_Review,Topological_atom_review_QiZhou_JPB2017}, have been employed to explore topological quantum matter in planar spaces \cite{TopologicalMatter_NPhy2016,Aidelsburger_review_2018,Topological_bands_coldatom_RevModPhys2019,topological_synthetic_dim_Ozawa_review2019}.
Material properties such as the topology of band structure 
\cite{Exp_Haldane_Nat2014,hexagonal_interferometer_Science2015, Chern_number_NatPhys_2015, ThoulessPumpBloch_NaturePhy2015, ThoulessPumpJapan_NaturePhy2016,hexagonal_Floquet_Science2016, hexagonal_Wilsonline_Science2016,Topology2DSOC_Science2016,microscopy_ladder_Nature_2017,SPT_fermion_SciAdv_2018,nodal_line_semimetal_NatPhys_2019}
can be probed with the high precision and tunability available in atomic systems.
On the other hand, there have been proposals for creating unconventional real spaces, such as a torus's surface \cite{lattice_torus_topology_PhysRevLett2018} 
or arbitrary Riemann surfaces \cite{hyperbolic_surfaces_Science_Bulletin_2021}, 
but relevant experiments remain elusive due to potential technical challenges in real space.


Synthetic spaces incorporating synthetic dimensions 
can bypass many constraints in the real space, holding promise for creating novel geometries with arbitrary dimensions.
Synthetic dimensions can be constructed using atomic internal states \cite{SytheticDimension_PhysRevLett2014, Quantum_simulation_nontrivial_topology_NJP2015}, momentum states \cite{SolitonSSH_NatCom2016,Topological_Anderson_insulator_Science2018}, 
time \cite{4D_quantum_Hall_Nature_2018}, or other degrees of freedom. 
Synthetic spaces have enabled experimental exploration of high-dimensional quantum matter, 
such as 4D quantum Hall systems \cite{4D_quantum_Hall_Nature_2018} and a Yang monopole in a 5D parameter space \cite{Yang_monopole_Science2018}.
Besides, manipulating boundary conditions is possible. This has allowed observations of edge states in synthetic 2D planes subject to magnetic fluxes
\cite{EdgeSate_Spielman_Science2015,EdgeSate_Italy_Science2015,Synthetic_dimension_SOC_clock_PRL_2016,SOC_OL_2017, large_Hall_strip_NatPhys_2020}. 
In addition, there have been rich ideas exploiting the versatile nature of synthetic spaces for creating unconventional spaces \cite{Fractional_QH_torus_PRA_2014,Quantum_simulation_nontrivial_topology_NJP2015,Interaction_PBC_NatCom_2015,Fractional_QH_real_cylinder_PRA_2016,Fractional_pump_synthetic_cylinder_PRL_2017,Yan_PRL_2018,Zhang_Hall_Tube_PhysRevA2020}, such as the surface of a cylinder, torus, or M\"obius strip, with or without gauge fields.
However, there remains very limited experimental exploration \cite{Hall_tube_PRL2019,Liang_tube_2020_PhysRevResearch2021} of topological quantum matter and transport in unconventional spaces.

Here, we realize a BEC on a synthetic cylindrical surface,
composed of a real spatial dimension and a curved synthetic dimension formed by cyclically coupled atomic spin states,
{subject to a net radial synthetic magnetic flux}.
We observe intriguing topological phenomena, {such as emergent topological band crossings and topological Bloch oscillations}, stemming from the interplay between gauge fields and nontrivial geometries of spaces.
{We also observe a topological transition manifested as gap opening at band crossings and further manipulate the band structure and BEC's transport via controlling the axial synthetic magnetic flux.
In striking contrast, these phenomena emerging on the Hall cylinder vanish when we unzip the cylinder into a planar Hall strip, illustrating the crucial and intriguing role of topology and geometries of spaces in novel topological quantum phenomena.}
 


\section{Experimental setup}
\begin{figure}[t!]
\includegraphics[width=3.4in]{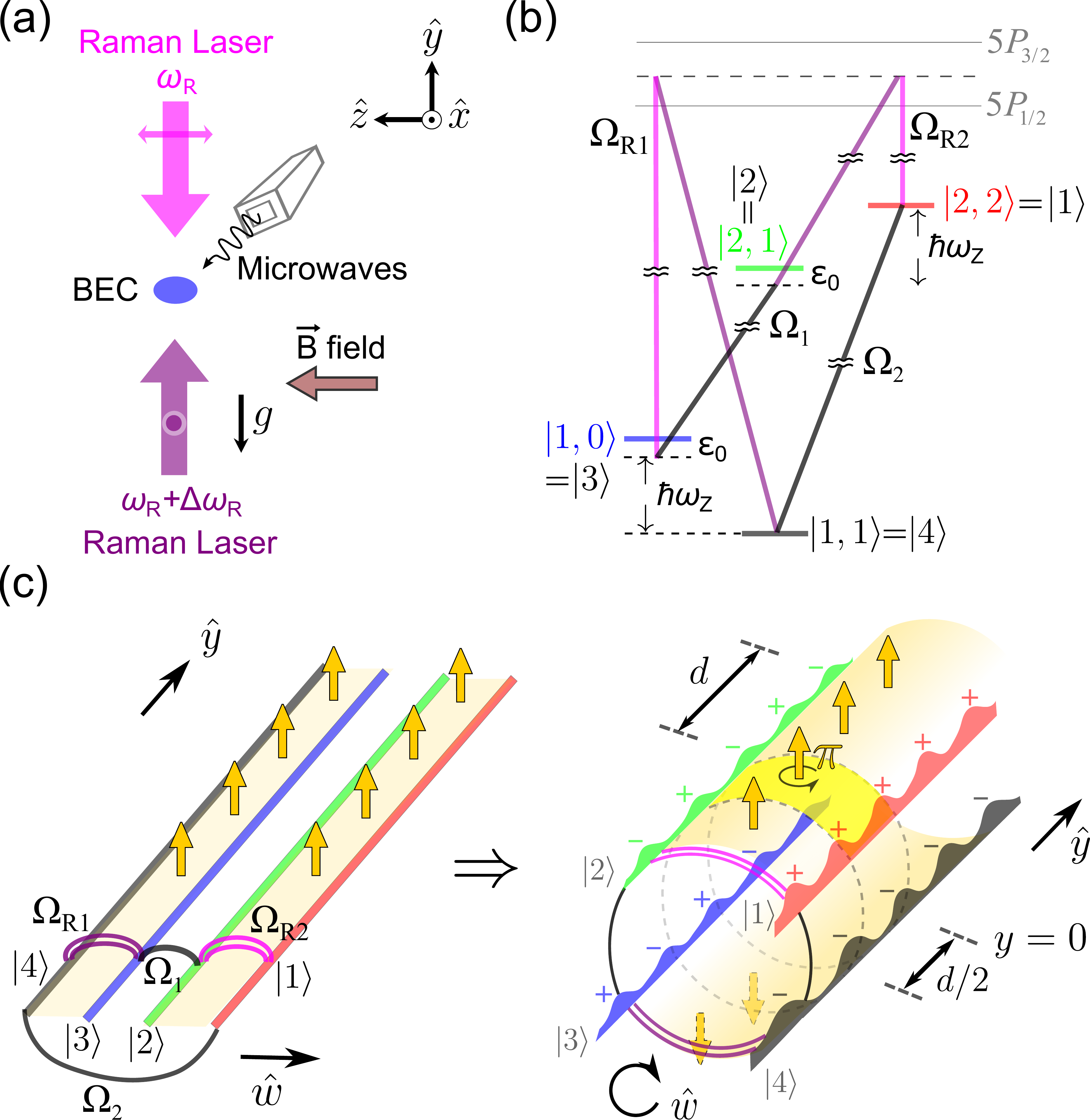}
\caption{\small{
Setup for realizing a synthetic Hall cylinder. 
(a) Counter-propagating Raman lasers with orthogonal linear polarizations (parallel to $\hat{z}$ and $\hat{x}$) and microwaves (propagating in the $x$-$y$ plane) are applied to a BEC with typical atom number (1-2)$\times 10^4$.
Gravity ``$g$'' is towards $-\hat{y}$. 
(b) Internal spin states $\left|1\right\rangle$, $\left|2\right\rangle$, $\left|3\right\rangle$, and $\left|4\right\rangle$ as discrete sites in the synthetic dimension $\hat{w}$ 
are cyclically coupled by Raman couplings $\Omega_\text{R1,R2}$ 
and microwave couplings $\Omega_{1, 2}$.
Linear Zeeman splitting $\hbar\omega_\text{Z}=\hbar\Delta\omega_\text{R}$ is generated by a bias magnetic field, 
where $\Delta \omega_\text{R}$ is the Raman lasers' angular frequency difference.
The quadratic Zeeman shift is $\varepsilon_0\approx 2.4$ $\E_r$. 
(c) Connecting two edges of a Hall strip (left) gives rise to a Hall cylinder (right), 
a cylindrical surface threaded by synthetic magnetic fields (yellow arrows in shaded regions).
The magnetic flux through a unit plaquette (highlighted area) formed by four maxima of the emergent BEC density modulations (wavy patterns with a periodicity of $d/2$; note that such density modulations do not occur in the Hall strip) corresponds to an Aharonov-Bohm phase of $\pi$.
The phase (with $+$ and $-$ respectively denoting $0$ and $\pi$ at positions of maximum density) of each spin component has a periodicity of either $d$ or $d/2$.
}}
\label{Fig1}
\end{figure}

In our experiments, a 
$^{87}$Rb BEC is produced in an optical dipole trap 
\cite{SpinE_NatCom_2019}. 
As shown in Figs.~\ref{Fig1}(a) and \ref{Fig1}(b), four internal spin states, 
$\left|F,m_F\right\rangle=\left|2,2\right\rangle, \left|2,1\right\rangle, \left|1,0\right\rangle, \left|1,1\right\rangle$, respectively relabeled as $\left|1\right\rangle, \left|2\right\rangle, \left|3\right\rangle, \left|4\right\rangle$, form discrete sites in the synthetic dimension 
(the $w$ direction), where $F$ ($m_F$) is the 
hyperfine spin (the magnetic quantum number). 
The synthetic dimension $w$ along with the real spatial dimension $y$ together span a synthetic cylindrical space.
Raman lasers 
along {$\pm\hat{y}$} couple $\left|1\right\rangle$ and $\left|2\right\rangle$ as well as $\left|3\right\rangle$ and $\left|4\right\rangle$ with respective Raman couplings $\Omega_\text{R2}$ and $\Omega_\text{R1}$. 
The Raman lasers' wavelength ($\lambda\approx790$ nm) defines
the photon recoil momentum $\hbar \k_r=2\pi\hbar/\lambda$ 
and recoil energy $\E_r=\hbar ^2\k_r^2/(2m)$ 
respectively used for 
units of momentum and energy, 
where $\hbar$ is the reduced Planck constant, and $m$ is the mass of $^{87}$Rb. 
Two microwave fields, with coupling strengths $\Omega_1$ and $\Omega_2$, 
respectively couple $\left|2\right\rangle$ and $\left|3\right\rangle$, and $\left|1\right\rangle$ and $\left|4\right\rangle$. 
This setup delivers a cyclic coupling (a periodic boundary condition) in the $w$ direction.
Differently from other cyclic couplings for creating 2D spin-orbit couplings \cite{Realistic_Rashba_SOC_PRA2011,2DSOC_Fermi2016} and Yang monopoles \cite{Yang_monopole_Science2018}, 
our scheme shown in Figs.~\ref{Fig1}(b) and \ref{Fig1}(c) connects two edges of a 2D planar Hall strip (with a synthetic magnetic field discussed below) and thus synthesizes the $y$ and the curved $w$ dimensions into a Hall cylinder.

For either the Hall strip or the Hall cylinder,
the synthetic magnetic flux is identical and engineered by making atoms obtain a phase after completing a closed trajectory on the $y$-$w$ surface.
An atom at location $y$ 
hopping along $\pm\hat{w}$ via a Raman transition
obtains a net momentum of $\hbar K$ along $\pm\hat{y}$, 
acquiring a Raman laser-imprinted $y$-dependent phase, $\pm K y$ (Appendix \ref{appendix::Hamiltonians}), where $K=2\k_r$. 
For the shaded regions in Fig.~\ref{Fig1}(c),
the phase acquired by an atom after traveling around an area of $\Delta y$ times one unit length along $\hat{w}$ is $\pm K\Delta y$.
Such a phase is analogous to the Aharonov-Bohm phase acquired by charged particles in a magnetic field, and thus corresponds to an artificial magnetic field and flux (yellow arrows) \cite{Harper_Ketterle_PhysRevLett2013,Harper_Bloch_PhysRevLett2013,SytheticDimension_PhysRevLett2014,EdgeSate_Spielman_Science2015,EdgeSate_Italy_Science2015}.

Since 
the transverse $x$ and $z$ directions are decoupled from the cylinder,
the single-particle Hamiltonian for atoms on the Hall cylinder [Fig.~\ref{Fig1}(c)]
is written in the basis of $\{\left|1\right\rangle, \left|2\right\rangle, \left|3\right\rangle, \left|4\right\rangle\}$ as (Appendix \ref{appendix::Hamiltonians})
\begin{eqnarray}
H&&=\frac{\hat{p}_y^2}{2m}\text{I}\nonumber\\
&&+\begin{pmatrix}
0 & \frac{\Omega_\text{R2}}{2}e^{-iK y}& 0 & \frac{\Omega_{2}}{2}\\ 
\frac{\Omega_\text{R2}}{2}e^{iK y} & \varepsilon_0 & \frac{\Omega_{1}}{2} & 0 \\
0 & \frac{\Omega_{1}}{2} & \varepsilon_0 & \frac{\Omega_\text{R1}}{2}e^{-iK y}\\
\frac{\Omega_{2}}{2} & 0 & \frac{\Omega_\text{R1}}{2}e^{iK y} & 0
\end{pmatrix},
\label{H}
\end{eqnarray}
where $\hat{p}_y=-i\hbar\frac{\partial}{\partial y}$, 
$\text{I}$ is the identity matrix, and $\varepsilon_0$ is the quadratic Zeeman shift.
Here, the Raman-induced 
$y$-dependent phase factor,
$e^{\pm iKy}$, cannot be gauged away due to our implemented periodic boundary condition,
unlike open boundary conditions such as when $\Omega_2=0$ (Appendix \ref{appendix::Hamiltonians}). 
This makes $H$ possess a translational symmetry under a translation of $d= 2\pi/K$.
Furthermore, $H$ has 
a nonsymmorphic symmetry \cite{Topological_atom_review_QiZhou_JPB2017}: a translation of $d/2$ along $\hat{y}$ followed by a unitary transformation along $\hat{w}$, $\left|1\right\rangle\rightarrow \left|1\right\rangle$, $\left|2\right\rangle\rightarrow -\left|2\right\rangle$, $|3\rangle\rightarrow -|3\rangle$, $\left|4\right\rangle\rightarrow \left|4\right\rangle$.

\section{Emergence of BEC crystalline order and topological band structure}
Even in the absence of an external lattice, the cylindrical surface (with the periodic boundary condition in the $\hat{w}$ direction) along with the radial synthetic magnetic flux cause the BEC to develop an emergent periodic or crystalline order in the $y$ direction with an underlying nonsymmorphic symmetry (Appendix \ref{appendix::symmetries_H}).
As sketched in Fig.~\ref{Fig1}(c) (see also {Fig.~{\ref{FigSIBECwavefunction_Mobius}}} in Appendix \ref{appendix::realspace_wavefunction}),
the BEC density (squared amplitude of the wavefunction) 
along $\hat{y}$ has a period of $d/2$ (half the period of $H$) while the phase of the BEC wavefunction of each spin component modulates with a period of either $d$ or $d/2$.
A plaquette [highlighted in Fig.~\ref{Fig1}(c)] formed by four maxima of the density modulations
thus has a magnetic flux $\Phi/\Phi_0= (Kd/2)/(2\pi)=1/2$ corresponding to a phase of $\pi$,
where $\Phi_0=2\pi\hbar/e$ is the magnetic flux quantum with $q\equiv-e$ ($e$ is the elementary charge) defined as the effective charge of a particle. 
Interestingly, here the emergent crystalline order corresponds to the generation of a subwavelength lattice having a periodicity of $d/2=\lambda/4$ 
(Anderson \textit{et al.}~{\cite{subwavelength_lattice_PhysRevResearch2020}} reported a more detailed study on this context).

To gain further insights, Fig.~\ref{Fig2}(a) illustrates 
how states in momentum space are coupled by the light fields.
These states are simply basis states for a generic lattice. 
However, they are coupled in a special way such that novel topological band structures can occur:
there are two independent branches (solid and dashed circles, each branch connected by lines representing couplings), with Hamiltonians $H_{i=1,2}(q_y)$, 
offset from each other by $K$, i.e.~$H_1(q_y)=H_2(q_y\pm K)$,
where $\hbar q_y$ is the quasimomentum.
Besides, each branch is invariant under a $2K$ translation, i.e.~$H_{i}(q_y)=H_{i}(q_y+n\times2K)$ where $n$ is an integer, thus corresponding to 
the $d/2$ periodicity in the BEC density modulations.
These two branches correspond to two groups of Bloch bands 
(each has a periodicity of $2K$) that are also offset from each other by $K$ and thus intersect at points occurring periodically by $K$ (Appendix \ref{appendix::symmetries_H}).
As shown in Fig.~\ref{Fig2}(b), the band structure possesses topological band crossing points that result from and are protected by the nonsymmorphic symmetry. 
Such band crossings are topologically robust against perturbations respecting the symmetry, such as variations of parameters in Equation~(\ref{H}), 
and have played important roles in topological quantum matter such as topological semimetals 
\cite{Topological_semimetal_review_NMaterial2016,Mobius_twist_PhysRevB2015,Topological_atom_review_QiZhou_JPB2017}.

\section{Topological band crossings and topological Bloch oscillations}
\begin{figure*}[t!]
\includegraphics[width=7.0in]{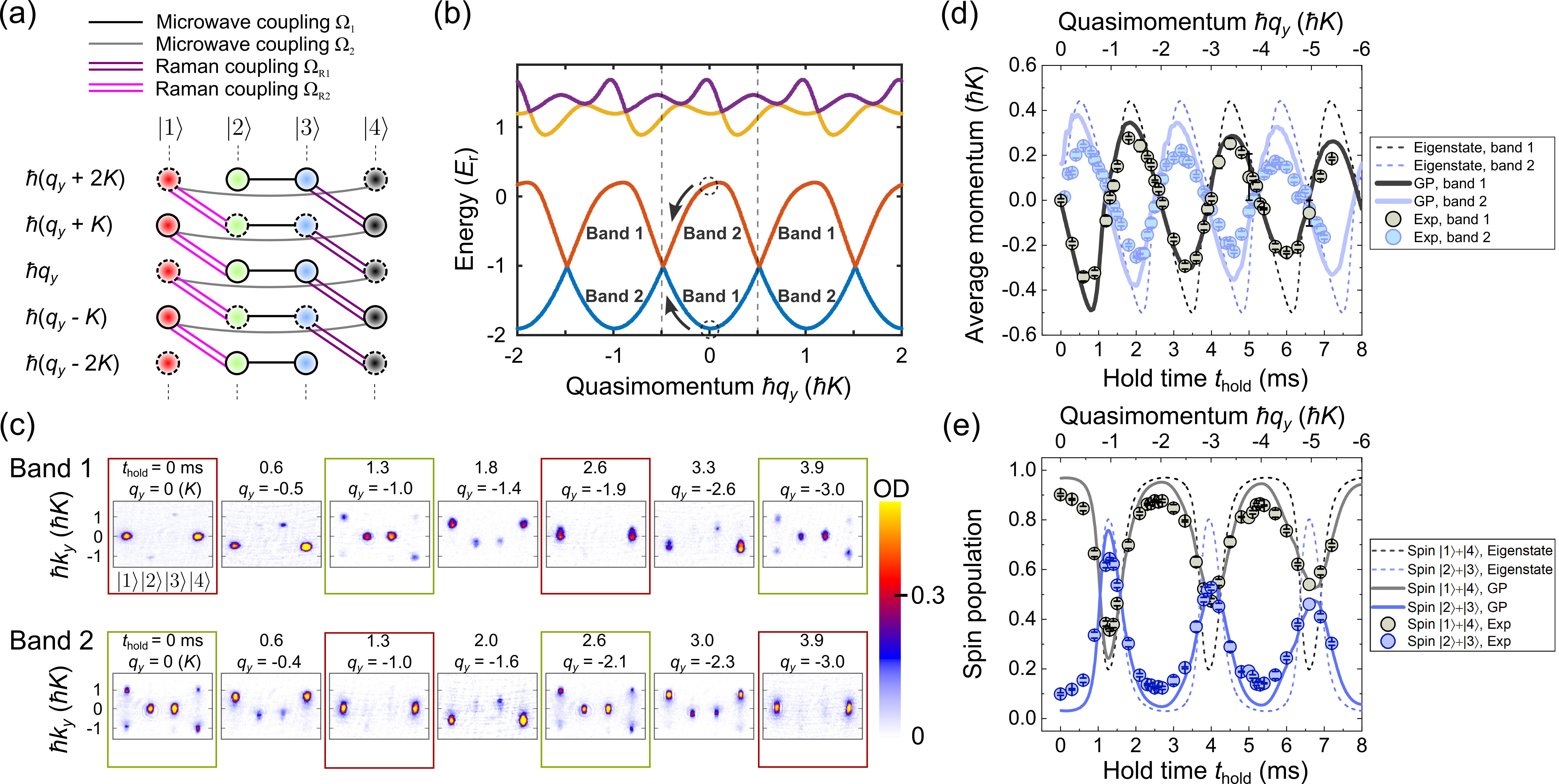}
\caption{\small{
Topological band crossings protected by the nonsymmorphic symmetry and observed topological Bloch oscillations. (a) In momentum space, basis states for a lattice are coupled by lasers and microwaves to form two independent branches (line-connected solid and dashed circles) offset from each other by $\hbar K$, manifesting the underlying nonsymmorphic symmetry. 
(b) Band structure (with a periodicity of $\hbar K$ in quasimomentum space) showing topological band crossings is calculated (see Appendix \ref{appendix::eigenstate_calculation}) using $\Omega_\text{R1(R2)}=-2.3 (3.2)$, $\Omega_{1(2)}=2.3 (3.3)$, and $\varepsilon_0=2.4$, all in units of $\E_r$. The first Brillouin zone is between the dashed lines. A BEC initially prepared around $q_y=0$ (dashed circles) in either band 1 or band 2 undergoes gravity-induced transport (indicated by arrows) and Bloch oscillations for various $\t_hold$. 
(c) Select TOF images showing atoms' spin and mechanical momentum compositions at various $\t_hold$ and the corresponding quasimomentum. Labels ``band 1'' and ``band 2'' imply different initial preparations. OD is the optical density.
(d) Average mechanical momentum of atoms versus $\t_hold$ and quasimomentum.
(e) (Fractional) spin population versus $\t_hold$ and quasimomentum for the transport in band 1. 
In (d, e), circles are experimental data (error bars are standard errors of typically 5 repetitive measurements).
Dashed lines (labeled as ``eigenstate'') are eigenstate calculations and solid lines (labeled as ``GP'') are Gross-Pitaevskii (GP) simulations; see
the text. These conventions also apply to Figs.~\ref{Fig3}(f),
\ref{Fig3}(g), and \ref{Fig4}(d).
}}
\label{Fig2}
\end{figure*}

To probe the band structure,
we perform spin- and momentum-resolved quantum transport measurements.
A BEC is initially prepared (Appendix \ref{appendix::initialstate_Exp}) around $q_y=0$ in either band 1 or band 2 [Fig.~\ref{Fig2}(b)].
Then, the dipole trap is abruptly turned off, allowing atoms to fall under gravity (towards $-\hat{y}$) for various holding times, 
$\t_hold$, during which the Raman and microwave couplings remain on.
In other words, the gravity induces transport of the BEC towards negative $q_y$ in Fig.~\ref{Fig2}(b) for various $\t_hold$.
Subsequently, we immediately turn off all 
coupling fields to release the atoms for a 15-ms time of flight (TOF) 
including a 9-ms spin-resolved Stern-Gerlach process, and then perform absorption imaging.
These TOF images unveil the mechanical momentum (along $\hat{y}$) and spin compositions of the atoms. 
We obtain atoms' average momentum by summing over all population-weighted mechanical momentum components (Appendix \ref{appendix::Imaging_analysis}).
The (fractional) population of a spin component is obtained by summing over (fractional) populations of all mechanical momentum components corresponding to that spin state. 

Figure \ref{Fig2}(c) shows select TOF images at various $\t_hold$ and 
the corresponding quasimomentum (the relation between $\t_hold$ and quasimomentum is determined by experimental calibration, see Fig.~{\ref{FigSI_qyvstime_calibration}} in Appendix \ref{appendix::calibration_quasimom_vs_t}).
The extracted average momentum and spin population are shown as circles in Figs.~\ref{Fig2}(d) and \ref{Fig2}(e), respectively. 
In Fig.~\ref{Fig2}(c), TOF images  show the reoccurrence of similar patterns (indicated by colored rectangles) with a period of $2\hbar K$ in quasimomentum ($2.6$ ms in time), 
consistent with the periodicity revealed in Figs.~\ref{Fig2}(d) and \ref{Fig2}(e).
These are Bloch oscillations \cite{BlochOsc_PhysRevLett1996} that possess twice the periodicity of the band structure ($\hbar K$),
analogous to traveling on a ``momentum-space M\"obius strip'': 
atoms have to travel twice the period of the band structure to reach the same quantum state, 
because at a band touching point they undergo a diabatic transition from the ground to the first excited bands.
Such period-multiplied topological Bloch oscillations
can unveil the band topology, 
which is characterized by a symmetry-protected topological invariant,
the period multiplier $\mu$ \cite{topological_bloch_osc_PRB2018, NonAbelian_Bloch_oscillations_NatCom2020} .
Here, the observed $2\hbar K$ periodicity ($\mu=2$) is protected by the nonsymmorphic symmetry, and is consistent with the $d/2$ periodicity of the density modulations [Fig.~\ref{Fig1}(c)]. 
Besides,
similar TOF images for band 1 and band 2 are offset from each other by $\hbar K$ or $1.3$ ms, 
consistent with the out-of-phase Bloch oscillations in Fig.~\ref{Fig2}(d).
These observed transport properties uncover the band crossings.

Unlike previous works using external optical lattices~\cite{SytheticDimension_PhysRevLett2014,EdgeSate_Spielman_Science2015,EdgeSate_Italy_Science2015,Hall_tube_PRL2019},
here, the emergent BEC crystalline order and topological band structure result from ``curving'' the Hall strip into the Hall cylinder and vanish on the Hall strip (which we realize by setting one of the microwave couplings to zero, see Fig.~{\ref{FigSI_Hall_Strip}} in Appendix \ref{appendix::HallStrip}).
Note that both the cylindrical surface and the net radial magnetic flux 
are essential for the emergent phenomena here, which disappear when either ingredient is absent (Appendix \ref{appendix::radialflux}). 

\begin{figure*}[t!]
\includegraphics[width=7.0in]{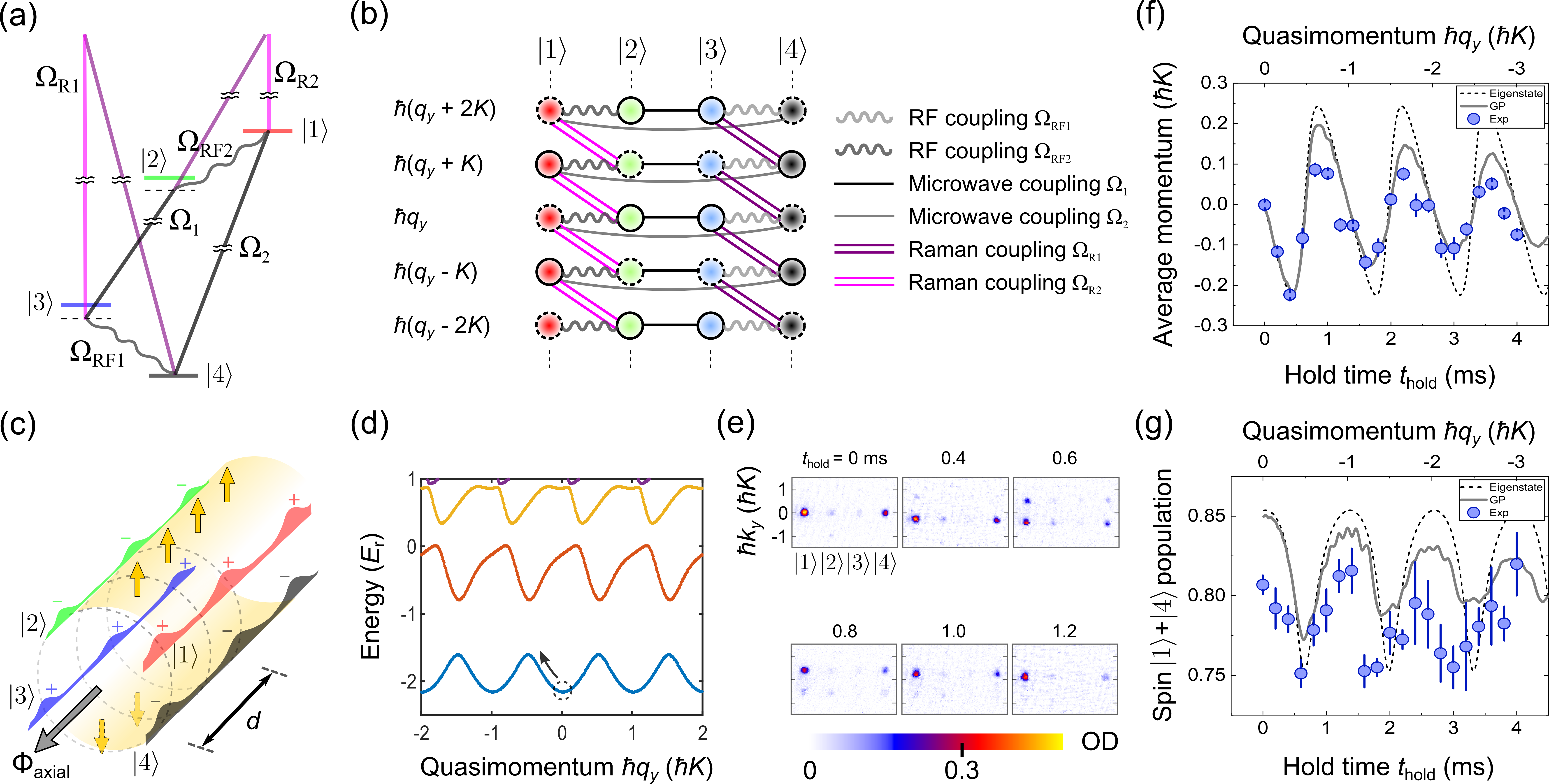}
\caption{\small{
Topological transition induced by breaking the nonsymmorphic symmetry and effects on transport.
(a) An RF wave matching the Raman lasers' frequency difference is applied to couple $\ket{1}$ and $\ket{2}$, and $\ket{3}$ and $\ket{4}$, with respective coupling strengths $\Omega_\text{RF2}$ and $\Omega_\text{RF1}$. 
(b) RF couplings (wiggling lines) merge the two independent branches in Fig.~\ref{Fig2}(a) into one, breaking the nonsymmorphic symmetry. 
(c) A cylinder with broken nonsymmorphic symmetry.
Both the BEC density and phase modulations have a periodicity of $d$ (see also Fig.~\ref{FigSIBECwavefunction_regular} in Appendix \ref{appendix::realspace_wavefunction}).
The synthetic axial magnetic field, represented by the thick black arrow, corresponds to a magnetic flux $\Phi_\text{axial}$ (see the text).
(d) Band structure showing band gap opening is calculated using similar parameters to those for Fig.~\ref{Fig2}(b) with additional RF couplings $\Omega_\text{RF1(RF2)}=1.4(-2.0)$ $\E_r$ with $\Phi_\text{axial}/\Phi_0=0.6$.
(e) Select TOF images at various $\t_hold$ and quasimomentum for transport of a BEC starting around $q_y=0$ [dashed circle in (d)] in the ground band.
Panels (f) and (g) respectively show atoms' average mechanical momentum and spin population versus $\t_hold$ and quasimomentum.
}}
\label{Fig3}
\end{figure*}

We have performed calculations using similar experimental parameters to gain further insights into experimental observations.
The calculation results are shown in Figs.~\ref{Fig2}(d) and \ref{Fig2}(e), as explained below.
First, we have calculated the average momentum and spin populations of the eigenstates of band 1 and band 2.
Results of such eigenstate calculations (dashed lines)
correspond to the transport of a noninteracting BEC.
We have also solved the time-dependent 3D Gross-Pitaevskii (GP) equation for an interacting BEC (Appendix \ref{appendix::GP}), showing results as solid lines.
The experimental results exhibit notable damping, consistent with GP simulations, but deviating from the eigenstate results that exhibit no damping.
These results show that inter-particle interactions, which can broaden the momentum distribution of atoms, are important {and can} cause damping of the transport and Bloch oscillations (see {Fig.~{\ref{FigSIGPcomparison}}} in Appendix \ref{appendix::GP}).

\section{Topological transition and effects on transport}
We can further break the nonsymmorphic symmetry that protects the band crossings 
by introducing a symmetry-breaking perturbation,
a radio frequency (RF) wave coupling
$\left|1\right\rangle$ and $\left|2\right\rangle$ as well as $\left|3\right\rangle$ and $\left|4\right\rangle$ 
with respective coupling strengths $\Omega_{\text{RF2}}$ and $\Omega_{\text{RF1}}$ [Fig.~\ref{Fig3}(a)].
This perturbation induces a topological transition manifested as gap opening at band crossings 
and makes the band structure now depend on a synthetic axial magnetic flux $\Phi_\text{axial}$ (no planar analogue) through the cylinder, as explained below.
In the momentum space, such an RF wave changes how the basis states for a lattice are coupled. 
RF couplings merge the two independent branches in Fig.~\ref{Fig2}(a) into one that has a $\hbar K$ periodicity [Fig.~\ref{Fig3}(b)], 
breaking the nonsymmorphic symmetry (Appendix \ref{appendix::symmetries_Hprime}).
Besides, any two basis states are now connected by multiple pathways, distinct from Fig.~\ref{Fig2}(a) in which there is either one or zero pathways for any two states in the same or different branches, respectively. 
These multiple pathways cause an interference effect controlled by 
an axial phase $\theta_\text{axial}$, 
where $\theta_\text{axial}=2\theta_{\text{RF}}+\theta_1-\theta_2$ (Appendix \ref{appendix::PhasesAxialFlux}), 
$\theta_{\text{RF}}$ is the phase associated with $\Omega_{\text{RF1,RF2}}$ {(generated from one RF wave, thus sharing the same RF phase)}, 
and $\theta_{1,2}$ is the phase associated with $\Omega_{1,2}$.
This axial phase $\theta_\text{axial}$ gives rise to $\Phi_\text{axial}$ [$\Phi_\text{axial}/\Phi_0=\theta_\text{axial}/2\pi$, Fig.~\ref{Fig3}(c)], 
which can affect the band structure as well as transport properties.
We can calibrate and precisely control $\Phi_\text{axial}$ by performing quench experiments (see Fig.~\ref{FigSI_phase_calibration} in Appendix \ref{appendix::calibration_axial_flux}).

To demonstrate band gap opening between the two lowest bands,
a specific band structure with a relatively large gap induced by a large RF coupling is chosen [Fig.~\ref{Fig3}(d)]
and probed by the same type of quantum transport measurement, 
in which a BEC is initially prepared at the bottom of the ground band. 
Figure~\ref{Fig3}(e) presents select TOF images at various $\t_hold$ and the corresponding quasimomentum. 
The extracted average momentum and spin population are shown as data points in Figs.~{\ref{Fig3}}(f) and {\ref{Fig3}}(g), respectively.
We observe that both momentum and spin oscillations exhibit half the period of those observed in Fig.~{\ref{Fig2}}.
This directly indicates that most atoms stay in the ground band (i.e., the transport is nearly adiabatic) because of the large gap opened.
That is, upon breaking the symmetry, the topological invariant $\mu$ changes from 2 to 1, 
signifying the topological transition manifested by the gap opening.
The observed Bloch oscillations in Figs.~{\ref{Fig3}}(f) and {\ref{Fig3}}(g) possess damping, 
again consistent with GP simulations (solid lines) that include effects of inter-particle interactions, but deviating from the non-interacting eigenstate calculations (dashed lines).

\section{Controlling transport via tuning axial synthetic magnetic flux}
\begin{figure}[t!]
\includegraphics[width=3.4in]{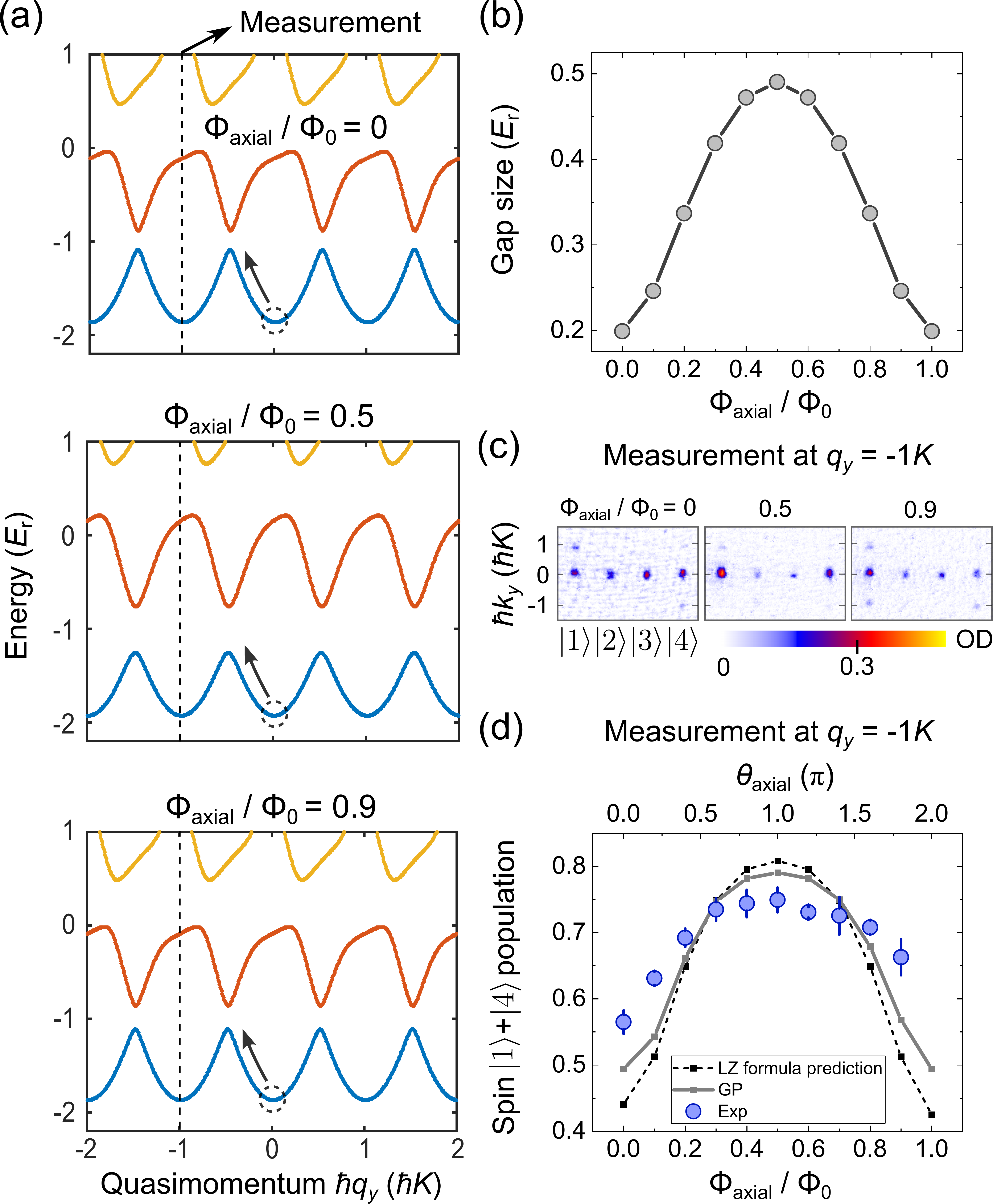}
\caption{\small{
Effects of the synthetic axial magnetic flux $\Phi_\text{axial}$ on band structures and transport.
(a) Band structures calculated using similar parameters to those for Fig.~\ref{Fig2}(b) 
with additional RF couplings $\Omega_\text{RF1(RF2)}=0.8(-1.1)$ $\E_r$ at representative values of $\Phi_\text{axial}$. 
We focus on transport measurement at $q_y=-1K$ 
(at $\t_hold=1.3$ ms, indicated by dashed lines) 
for a BEC initially prepared around $q_y=0$ (dashed circles).
(b) Calculated band gap size between the two lowest bands at various $\Phi_\text{axial}$. 
(c) Select TOF images measured at $q_y=-1K$ with representative values of $\Phi_\text{axial}$. 
(d) Spin population versus $\Phi_\text{axial}$ measured at $q_y=-1K$, compared with GP simulations (solid line) and the prediction (dashed line) by the Landau-Zener formula; see the text.
}}
\label{Fig4}
\end{figure}

Lastly, we demonstrate that controlling $\Phi_\text{axial}$ allows manipulating both the band structure and transport properties.
Whereas this axial flux can be gauged away when the nonsymmorphic symmetry protects the band crossings, it becomes important when RF couplings lift the symmetry and open the band gap, due to the previously discussed interference between pathways in the momentum space (see details in Appendix \ref{appendix::PhasesAxialFlux}).
As shown in Fig.~\ref{Fig4}(a), we perform the same type of quantum transport 
with a BEC initially prepared around $q_y=0$ in band structures at moderate RF 
couplings (giving an intermediate gap size) with various $\Phi_\text{axial}$,
focusing on measuring spin populations at $q_y=-1K$ 
(at $\t_hold=1.3$ ms, indicated by dashed lines).
Such spin populations are sensitive to the $\Phi_\text{axial}$-dependent gap size [Fig.~\ref{Fig4}(b)] between the two lowest bands, 
because different gap sizes can cause notable differences in atoms' Landau-Zener tunneling probability \cite{Olson_LZ_PhysRevA2014} to the excited band.
Figure~\ref{Fig4}(c) presents select TOF images for the measurements at $q_y=-1K$ with various $\Phi_\text{axial}$. 
The extracted spin population at various $\Phi_\text{axial}$, shown in Fig.~\ref{Fig4}(d), 
is consistent with both the GP simulations (solid lines) 
and the prediction (dashed lines) by the Landau-Zener formula (considering the gap size and eigenspin compositions of the two lowest bands at various $\Phi_\text{axial}$).  
The observed strong dependence of a transport property (here spin population following transport) on the synthetic axial magnetic flux, 
reminiscent of a ``magnetotransport'' behavior {{\cite{Luis_Natnano_2016}}}, 
{reflects} the underlying $\Phi_\text{axial}$-dependent band structures.


\section{Discussion}
Our work differs from another Hall cylinder \cite{Hall_tube_PRL2019} recently realized in a number of fundamental ways.
(1) Our setup does not use an external real-space optical lattice as in Ref.~\cite{Hall_tube_PRL2019} 
(we also noted another recent experiment \cite{Liang_tube_2020_PhysRevResearch2021} exploring incommensurability-induced effects, where an external optical lattice is important).
Rather, BEC crystalline order and topological band structure emerge due to curving the Hall strip into a Hall cylinder and thus are intrinsic properties of a Hall cylinder, not relying on an extra real-space lattice potential.
(2) Topological band structures are distinct. We observe topological band crossings protected by a nonsymmorphic symmetry, 
while Ref.~\cite{Hall_tube_PRL2019} revealed a gapped topological band protected by a generalized inversion symmetry.
Based on (1) and (2), 
our work and Ref.~{\cite{Hall_tube_PRL2019}} respectively reveal the intrinsic and extrinsic topological properties of a Hall cylinder. 
Besides, once the emergent intrinsic crystalline order is incommensurate with the external lattice, intriguing localization phenomena are predicted to arise {\cite{Localization_Hall_Cylinder_PhysRevLett2021}},
where the axial flux would strongly influence the physical observables when near the localization-delocalization transitions.
In fact, all these phenomena reflect the rich physics arising on a Hall cylinder yet absent in the planar counterpart (the Hall strip). 
This illustrates the crucial and intriguing role of geometries of spaces in novel topological phenomena.
(3) We perform quantum transport measurement to probe the band structures 
as well as the topological transition and demonstrate topological Bloch oscillations.
We also conduct quench experiments to calibrate the axial synthetic magnetic flux.
In Ref.~\cite{Hall_tube_PRL2019}, quench dynamics is used to probe a band gap closing associated with a topological transition {from trivial to topological gapped bands, where the transition is induced by varying one of the couplings between spin states}.
(4) We demonstrate the capability of tuning the axial synthetic magnetic flux to control the band structure as well as BEC's transport. 
References \cite{Zhang_Hall_Tube_PhysRevA2020, Liang_tube_2020_PhysRevResearch2021,Localization_Hall_Cylinder_PhysRevLett2021} point out that such a flux should play an important role but it is not considered in Ref.~\cite{Hall_tube_PRL2019}.

\section{Conclusions and outlook}
In summary, engineering the geometry of space subject to a synthetic magnetic field has allowed us to
observe an emergent topological band structure and a topological transition as probed by quantum transport measurements.
Our work may offer valuable insights to exploring novel quantum matter intrinsic to unconventional spaces, a multidisciplinary direction in quantum science and engineering 
that is of high interest to broad communities of atomic and molecular physics, condensed matter physics, and photonics quantum simulations {\cite{topological_synthetic_dim_Ozawa_review2019}}.
Future directions may include investigating topological transitions induced by the tunable axial magnetic flux \cite{Zhang_Hall_Tube_PhysRevA2020}, 
studying superfluidity on curved surfaces \cite{Cylinder_Ho_PhysRevLett2015,Vortex_cylindrical_PRA2017}, 
implementing a Laughlin's charge pump \cite{Laughlin_pump_PhysRevB1981} [e.g., by making Equation~(\ref{H}) time dependent], 
exploring the fractal energy spectrum of Hofstadter's butterfly as suggested in Ref.~\cite{SytheticDimension_PhysRevLett2014},
and using Laguerre-Gaussian beams for the Raman lasers to create a Hall torus \cite{Yan_PRL_2018}. 
Even more possibilities arise if inter-particle interactions can be tuned by means such as optical lattices \cite{Bloch_RevModPhys2008} or Feshbach resonances \cite{Feshbach_resonances_RevModPhys2010}.
For example, it is interesting to study quantum many-body phases such as
topologically ordered states, for example fractional quantum Hall-like
states, on a Hall cylinder or torus 
\cite{Fractional_QH_torus_PRA_2014,Interaction_PBC_NatCom_2015,Fractional_QH_real_cylinder_PRA_2016, Fractional_pump_synthetic_cylinder_PRL_2017, large_Hall_strip_NatPhys_2020} or in curved spaces such as hyperbolic surfaces \cite{hyperbolic_surfaces_Science_Bulletin_2021}.

Future experiments can also study how interactions can affect the quantum transport, 
such as Landau-Zener transitions \cite{LZ_lattice_PRA2002,ManybodyLZ_NatPhy2011,Hysteresis_superfluid_ring_Nat2014}.
Our GP simulations (Fig.~\ref{FigSI_LZ} in Appendix \ref{appendix::effects_interactions_LZ}) performed for a BEC on a synthetic Hall cylinder 
(but in a 1D trap different from our current experimental setup to enhance the interaction effects) 
have revealed that the Landau-Zener tunneling from ground to excited bands 
increases and approaches 1 with increasing interactions.
This suggests that strong interactions could make the quantum transport of the BEC more diabatic, effectively experiencing a different topology (M\"obius strip) in the momentum space, compared to a non-interacting BEC.

Our technique can be extended to engineer other interesting geometries {\cite{Quantum_simulation_nontrivial_topology_NJP2015}} subject to gauge fields,
such as a Hall torus by using Laguerre-Gaussian beams for the Raman lasers {\cite{Yan_PRL_2018}} 
or two Hall tori (or cylinders) glued together via additional couplings {\cite{Yan_PRL_2018}}. 
Such unconventional spaces may be relatively challenging to fabricate using external optical lattices or in other conventional systems. 
Therefore, our technique may shed new light on engineering synthetic spaces of nontrivial geometries subject to gauge fields.

\begin{acknowledgments}
Our experiment has been supported by NSF grants PHY-1708134 and PHY-2012185. Our theoretical work has been supported in part by NSF through Grant No.~PHY-1806796 and the
Air Force Office of Scientific Research under Grant No.~FA9550-20-1-0221.
C.H.L. thank Esat H.~Kondakci and Su-Ju Wang for helpful discussions. D.B.B. also acknowledges support by the Purdue Research Foundation Ph.D. fellowship.
C.H.L., Q.Z. and Y.P.C. also acknowledge
support from DOE, Office of Science through the Quantum
Science Center (QSC), a National Quantum Information Science Research Center during the final phase of this work. 
\end{acknowledgments}

\appendix

\section{\label{appendix::Hamiltonians}Single-particle Hamiltonians}
To derive the Hamiltonians, here we relabel the spin states $\left|2,2\right\rangle=\left|\tilde{1}\right\rangle$, $\left|2,1\right\rangle=\left|\tilde{2}\right\rangle$, $\left|1,0\right\rangle=\left|\tilde{3}\right\rangle$, and $\left|1,1\right\rangle=\left|\tilde{4}\right\rangle$ (where the tilde refers to a non-rotating frame, as explained below), with respective energies $E_1$, $E_2$, $E_3$, and $E_4$.
Raman lasers along $\pm \hat{y}$ with an angular frequency difference $\Delta\omega_\text{R}$ couple $\left|\tilde{1}\right\rangle$ and $\left|\tilde{2}\right\rangle$, and $\left|\tilde{3}\right\rangle$ and $\left|\tilde{4}\right\rangle$, 
with respective coupling strengths $\Omega_\text{R2}$ and $\Omega_\text{R1}$. 
Two microwaves with angular frequencies $\omega_{1}$ and $\omega_{2}$ couple $\left|\tilde{2}\right\rangle$ and $\left|\tilde{3}\right\rangle$, and $\left|\tilde{1}\right\rangle$ and $\left|\tilde{4}\right\rangle$, with respective coupling strengths $\Omega_{1}$ and $\Omega_{2}$.
Note that {$|\Omega_\text{R1}|\neq|\Omega_\text{R2}|$ ($\Omega_\text{R1}<0$, 
$\Omega_\text{R2}>0$)} and $\Omega_{1}\neq\Omega_{2}$ {($\Omega_{1,2}>0$)}
because of the different Clebsch-Gordan coefficients associated with different atomic transitions.


We define $E_3-E_4=\hbar\omega_\text{Z}+\varepsilon_0$ and $E_1-E_2=\hbar\omega_\text{Z}-\varepsilon_0$, where $\hbar\omega_\text{Z}$ is the effective linear Zeeman splitting and $\varepsilon_0$ is the effective quadratic Zeeman shift. 
In our experiments, $\varepsilon_0\sim2.4$ $E_r$, given by the applied bias magnetic field (about 5 gauss). 
We define the (two-photon) Raman laser detuning $\delta_\text{R} = (2\hbar\Delta\omega_R-(E_3-E_4)-(E_1-E_2))/2 = \hbar(\Delta\omega_\text{R}-\omega_\text{Z})$
and the (one-photon) microwave detunings $\delta_{1} = \hbar\omega_{1}-(E_2-E_3)$ and $\delta_{2} = \hbar\omega_{2}-(E_1-E_4)$.

In the following, we derive the single-particle Hamiltonians for various coupling schemes: 
(1) the Hamiltonian $H$ 
and the corresponding momentum-space Hamiltonian $H_{q_y}$ 
for the Hall cylinder with a nonsymmorphic symmetry; (2) the Hamiltonian $H'$ and the corresponding momentum-space Hamiltonian $H'_{q_y}$ 
for the Hall cylinder with a broken nonsymmorphic symmetry; (3) the momentum-space Hamiltonian $H_{\text{strip}}$ for the Hall strip.

\subsection{Hall cylinder with the nonsymmorphic symmetry}
The free atomic Hamiltonian taking into account the motion along $\hat{y}$ is written as:
\begin{eqnarray}
\label{H_free}
\tilde{H}_{\text{free}} = &&\frac{\hat{p}_y^2}{2m}\text{I}+E_1\left|\tilde{1}\right\rangle\left\langle\tilde{1}\right|+E_2\left|\tilde{2}\right\rangle\left\langle\tilde{2}\right|+E_3\left|\tilde{3}\right\rangle\left\langle\tilde{3}\right|\nonumber\\
&&+E_4\left|\tilde{4}\right\rangle\left\langle\tilde{4}\right|,
\end{eqnarray}
where $\text{I}$ is the identity matrix 
and $\hat{p}_y=-i\hbar\frac{\partial}{\partial y}$ is the momentum operator along $\hat{y}$. 
In the rotating-wave approximation, the Hamiltonians describing the Raman ($\tilde{H}_{\text{Raman}}$) and microwave ($\tilde{H}_{1,2}$) couplings are respectively written as:
\begin{eqnarray}
\label{H_Raman}
\tilde{H}_{\text{Raman}} = &&e^{i(-K y-\Delta\omega_\text{R} t)}(\frac{\Omega_\text{R2}}{2}\left|\tilde{1}\right\rangle\left\langle\tilde{2}\right|+\frac{\Omega_\text{R1}}{2}\left|\tilde{3}\right\rangle\left\langle\tilde{4}\right|)\nonumber\\
&&+\text{H.c.},
\end{eqnarray}
\begin{align}
\label{H_M1}
\tilde{H}_{1} = \frac{\Omega_{1}}{2}e^{-i\omega_{1}t}(\left|\tilde{2}\right\rangle\left\langle\tilde{3}\right|)+\text{H.c.},
\end{align}
\begin{align}
\label{H_M2}
\tilde{H}_{2} = \frac{\Omega_{2}}{2}e^{-i\omega_{2}t}(\left|\tilde{1}\right\rangle\left\langle\tilde{4}\right|)+\text{H.c.},
\end{align}
where $K=2\k_r$, $\text{H.c.}$ stands for Hermitian conjugate, and the initial phases of these coupling fields are ignored temporarily but will be considered later. 

\vspace{2mm}
We choose a rotating frame defined by the following unitary transformations to eliminate the time-dependent terms in Eqs.~(\ref{H_Raman}-\ref{H_M2}): 
\begin{align}
\label{Eq_rotating_frame}
\left|\tilde{1}\right\rangle&=e^{i\Delta\omega_\text{R}t}\left|1\right\rangle e^{i\frac{\omega_{1}}{2}t}, 
\left|\tilde{2}\right\rangle= e^{i\frac{\omega_{1}}{2}t}\left|2\right\rangle,&&\\\nonumber
\left|\tilde{3}\right\rangle&= e^{-i\frac{\omega_{1}}{2}t}\left|3\right\rangle,
\left|\tilde{4}\right\rangle=e^{-i\Delta\omega_\text{R}t}\left|4\right\rangle e^{-i\frac{\omega_{1}}{2}t}.&&
\end{align}
In such a rotating frame (without tildes),
\label{H_free,rot}
\begin{eqnarray}
H_{\text{free}} = &&\frac{\hat{p}_y^2}{2m}\text{I}+(E_1-\hbar\Delta\omega_\text{R}-\frac{\hbar\omega_{1}}{2})\left|1\right\rangle\left\langle1\right|\nonumber\\
&&+(E_2-\frac{\hbar\omega_{1}}{2})\left|2\right\rangle\left\langle2\right|
+(E_3+\frac{\hbar\omega_{1}}{2})\left|3\right\rangle\left\langle3\right|\nonumber\\
&&+(E_4+\hbar\Delta\omega_\text{R}+\frac{\hbar\omega_{1}}{2})\left|4\right\rangle\left\langle4\right|
\end{eqnarray}
\begin{align}
\label{H_Raman_rot}
H_{\text{Raman}} = e^{-i K y}(\frac{\Omega_\text{R2}}{2}\left|1\right\rangle\left\langle 2\right|+\frac{\Omega_\text{R1}}{2}\left|3\right\rangle\left\langle4\right|)+\text{H.c.}
\end{align}
\begin{align}
\label{H_M1_rot}
H_{1} = \frac{\Omega_{1}}{2}(\left|2\right\rangle\left\langle3\right|)+\text{H.c.}
\end{align}
\begin{align}
\label{H_M2_rot}
H_{2} = \frac{\Omega_{2}}{2}e^{-i\omega_{2}t}e^{i2\Delta\omega_\text{R} t}e^{i\omega_{1}t}(\left|1\right\rangle\left\langle4\right|)+\text{H.c.}
\end{align}
where $H_{\text{Raman}}$ and $H_{1}$ become time independent. 
By further requiring
\begin{align}
{\omega_{2}}={2\Delta\omega_\text{R} }+{\omega_{1}}, 
\label{resonance_condition}
\end{align}
Eq.~(\ref{H_M2_rot}) becomes $H_{2} = \frac{\Omega_{2}}{2}(\left|1\right\rangle\left\langle4\right|)+\text{H.c.}$, which is also time independent. 
Eq.~(\ref{resonance_condition}) is called the resonance condition for the cyclic coupling and is realized in this work [as depicted in Fig.~\ref{Fig1}(b)].

Therefore, in the rotating frame defined by Eq.~(\ref{Eq_rotating_frame}) and when the resonance condition in Eq.~(\ref{resonance_condition}) is fulfilled, $H=H_{\text{free}}+H_{\text{Raman}}+H_{1}+H_{2}$ is time independent. 
In the basis of $\{\left|1\right\rangle, \left|2\right\rangle, \left|3\right\rangle, \left|4\right\rangle\}$,
\begin{widetext}
\begin{equation}
\label{H_realspace}
H=\frac{\hat{p}_y^2}{2m}\text{I}+
\begin{pmatrix}
E_1-\hbar\Delta\omegaR-\frac{\hbar\omega_{1}}{2} & \frac{\Omega_\text{R2}}{2}e^{-i K y} & 0 & \frac{\Omega_{2}}{2}\\ 
\frac{\Omega_\text{R2}}{2}e^{i K y} & E_2-\frac{\hbar\omega_{1}}{2} & \frac{\Omega_{1}}{2} & 0 \\
0 & \frac{\Omega_{1}}{2} & E_3+\frac{\hbar\omega_{1}}{2} & \frac{\Omega_\text{R1}}{2}e^{-i K y}\\
\frac{\Omega_{2}}{2} & 0 & \frac{\Omega_\text{R1}}{2}e^{i K y} & E_4+\hbar\Delta\omegaR+\frac{\hbar\omega_{1}}{2}
\end{pmatrix}.
\end{equation}
\end{widetext}

The above equation shows that a Raman transition corresponds to a $y$-dependent phase factor, $e^{\pm iK y}$, while a microwave transition does not lead to a position-dependent phase. 
Redefining all energies such that $E_3+\hbar\omega_1/2=\varepsilon_0+\delta_{1}/2$ and using the definitions of $\varepsilon_0$, $\deltaR$, $\delta_{1}$, $\delta_{2}$, and the resonance condition in Eq.~(\ref{resonance_condition}), we obtain 
\begin{align}
    \delta_{1}+2\deltaR = \delta_{2}
    \label{detuning_relation}
\end{align}
and rewrite Eq.~(\ref{H_realspace}) as
\begin{eqnarray}
\label{H_static_realspace}
&&H=\frac{\hat{p}_y^2}{2m}\text{I}\nonumber\\
&&+\begin{pmatrix}
-\deltaR-\frac{\delta_{1}}{2} & \frac{\Omega_\text{R2}}{2}e^{-i K y} & 0 & \frac{\Omega_{2}}{2}\\ 
\frac{\Omega_\text{R2}}{2}e^{i K y} & \varepsilon_0-\frac{\delta_{1}}{2} & \frac{\Omega_{1}}{2} & 0 \\
0 & \frac{\Omega_{1}}{2} & \varepsilon_0+\frac{\delta_{1}}{2} & \frac{\Omega_\text{R1}}{2}e^{-i K y}\\
\frac{\Omega_{2}}{2} & 0 & \frac{\Omega_\text{R1}}{2}e^{i K y} & \deltaR+\frac{\delta_{1}}{2}
\end{pmatrix}.
\end{eqnarray}
\vspace{6mm}
This equation includes Raman and microwave detunings, which can be nonzero during the initial state preparation process as discussed later. 
After the initial state preparation, $\deltaR=\delta_1=\delta_2=0$ is achieved 
and Eq.~(\ref{H_static_realspace}) becomes Eq.~(1).
Thus, all the detunings are zero in the main text.

To calculate band structures, we derive the momentum-space Hamiltonian $H_{q_y}$ by considering the coupling scheme in Fig.~\ref{Fig2}(a). 
The spin and mechanical momentum states comprise a plane-wave basis, denoted by
\begin{align}
\{\left|\hbar(q_y+ nK);m\right\rangle\}=\{e^{i(q_y+ nK)y}\ket{m}\},
\label{plane_wave_basis}
\end{align}
where $\hbar(q_y+ nK)$ is the mechanical momentum, $m=1,2,3,4$ label the spin, $\hbar q_y$ is the quasimomentum, and $n$ is an integer. Then, $H_{q_y}$ reads
\begin{align}
H_{q_y}=
\begin{pmatrix}
\ddots & \vdots & \vdots & \vdots & \vdots & \vdots & \reflectbox{$\ddots$} \\
\dots & A_{-2} & B & 0 & 0 & 0 & \dots\\ 
\dots & B^{\dagger} & A_{-1} & B & 0 & 0 & \dots\\
\dots & 0 & B^{\dagger} & A_{0} & B & 0 & \dots\\
\dots & 0 & 0 & B^{\dagger} & A_{1} & B & \dots\\
\dots & 0 & 0 & 0 & B^{\dagger} & A_{2} & \dots\\
\reflectbox{$\ddots$} & \vdots & \vdots & \vdots & \vdots & \vdots & \ddots
\end{pmatrix}
,
\label{H_Momentum}
\end{align}
where the $A_n$ matrices are on the diagonal of $H_{q_y}$. Each $A_n$ is a 4-by-4 matrix written in the basis of $\{\left|\hbar(q_y+nK);m\right\rangle\}$, where the four spin states have identical mechanical momentum (i.e., same $n$). Thus, $A_n$ only includes microwave couplings. 
When all the detunings are zero,
\onecolumngrid
\begin{equation}
\label{An_sameMomentum}
A_n=
\begin{pmatrix}
\frac{\hbar^2}{2m}(q_y+nK)^2 & 0 & 0 & \frac{\Omega_{2}}{2}\\ 
0 & \frac{\hbar^2}{2m}(q_y+nK)^2+\varepsilon_0 & \frac{\Omega_{1}}{2} & 0 \\
0 & \frac{\Omega_{1}}{2} & \frac{\hbar^2}{2m}(q_y+nK)^2+\varepsilon_0 & 0\\
\frac{\Omega_{2}}{2} & 0 & 0 & \frac{\hbar^2}{2m}(q_y+nK)^2
\end{pmatrix}.
\end{equation}
Here $B$ is a 4-by-4 matrix responsible for the Raman coupling between adjacent $A_n$ matrices:
\begin{align}
B=
\begin{pmatrix}
0 & 0 & 0 & 0\\
\Omega_\text{R2}/2 & 0 & 0 & 0\\
0 & 0 & 0 & 0\\
0 & 0 & \Omega_\text{R1}/2 & 0
\end{pmatrix}.
\label{Bmatrix}
\end{align}

\subsection{Hall cylinder with a broken nonsymmorphic symmetry}
An RF wave whose angular frequency equals $\Delta\omega_\text{R}$
couples $\ket{1}$ and $\ket{2}$, and $\ket{3}$ and $\ket{4}$, with respective coupling strengths $\Omega_\text{RF2}$ and $\Omega_\text{RF1}$.
Note that {$|\Omega_\text{RF1}|\neq|\Omega_\text{RF2}|$ ($\Omega_\text{RF1}>0$ and $\Omega_\text{RF2}<0$)} due to the different Clebsch-Gordan coefficients associated with different transitions.
The corresponding Hamiltonian $H'$ is obtained by adding the RF couplings to $H$ as
\begin{equation}
 H'
=\frac{\hat{p}_y^2}{2m}\text{I}+
\begin{pmatrix}
-\deltaR-\frac{\delta_{1}}{2} & \frac{\Omega_\text{RF2}}{2}+\frac{\Omega_\text{R2}}{2}e^{-iK y} & 0 & \frac{\Omega_{2}}{2}\\ 
\frac{\Omega_\text{RF2}}{2}+\frac{\Omega_\text{R2}}{2}e^{iK y} & \varepsilon_0-\frac{\delta_{1}}{2} & \frac{\Omega_{1}}{2} & 0 \\
0 & \frac{\Omega_{1}}{2} & \varepsilon_0+\frac{\delta_{1}}{2} & \frac{\Omega_\text{RF1}}{2}+\frac{\Omega_\text{R1}}{2}e^{-iK y}\\
\frac{\Omega_{2}}{2} & 0 & \frac{\Omega_\text{RF1}}{2}+\frac{\Omega_\text{R1}}{2}e^{iK y} & \deltaR+\frac{\delta_{1}}{2}
\end{pmatrix}.
\label{H'}   
\end{equation}
Let $\deltaR=\delta_1=\delta_2=0$ for simplicity.
Since the RF wave only couples spin states that have the same mechanical momentum,
the corresponding momentum-space Hamiltonian $H'_{q_y}$ has the same form as Eq.~(\ref{H_Momentum}) but with a modified $A_n$ denoted by $A_{n}'$:

\begin{equation}
    \label{An_RF}
A_{n}'=
\begin{pmatrix}
\frac{\hbar^2}{2m}(q+nK)^2 & \frac{\Omega_\text{RF2}}{2} & 0 & \frac{\Omega_{2}}{2}\\ 
\frac{\Omega_\text{RF2}}{2} & \frac{\hbar^2}{2m}(q+nK)^2+\varepsilon_0 & \frac{\Omega_{1}}{2} & 0 \\
0 & \frac{\Omega_{1}}{2} & \frac{\hbar^2}{2m}(q+nK)^2+\varepsilon_0 & \frac{\Omega_\text{RF1}}{2}\\
\frac{\Omega_{2}}{2} & 0 & \frac{\Omega_\text{RF1}}{2} & \frac{\hbar^2}{2m}(q+nK)^2
\end{pmatrix}.
\end{equation}

\subsection{Hall strip}
In this case, only $\Omega_\text{R1,R2}$ and $\Omega_{1}$ are present. 
To derive the corresponding Hamiltonian $H_\text{strip}$,
we apply a unitary transformation 
\begin{equation}\label{eq:HallStripUnitary}
\hat{U_0}=\left(
\begin{array}{cccc}
 e^{i K y} & 0 & 0 & 0 \\
 0 & 1 & 0 & 0 \\
 0 & 0 & 1 & 0 \\
 0 & 0 & 0 & e^{-i K y} \\
\end{array}
\right)
\end{equation}
to $H$ in Eq.~({\ref{H_static_realspace}}), $\hat{U_0}H\hat{U_0}^{-1}$,  with $\Omega_2=0$ and $\deltaR=\delta_1=\delta_2=0$.
Noting that $\hat{p}_y^2/(2m) \text{I}=\hbar^2q_y^2/(2m) \text{I}$ for the plane-wave basis, this transformation gauges away the $y$-dependent phase factor in $H$
and leads to $H_\text{strip}$. 
In the basis of $\{\left|1\right\rangle, \left|2\right\rangle, \left|3\right\rangle, \left|4\right\rangle\}$,

\begin{equation}
    \label{H_unzipped}
H_\text{strip}=
\begin{pmatrix}
\frac{\hbar^2}{2m}(q_y+K)^2 & \frac{\Omega_\text{R2}}{2} & 0 & 0\\ 
\frac{\Omega_\text{R2}}{2} & \frac{\hbar^2}{2m}(q_y)^2+\varepsilon_0 & \frac{\Omega_{1}}{2} & 0 \\
0 & \frac{\Omega_{1}}{2} & \frac{\hbar^2}{2m}(q_y)^2+\varepsilon_0 & \frac{\Omega_\text{R1}}{2}\\
0 & 0 & \frac{\Omega_\text{R1}}{2} & \frac{\hbar^2}{2m}(q_y-K)^2
\end{pmatrix}.
\end{equation}
It is important to realize that if $\Omega_2\neq 0$, the transformation 
cannot gauge away the $y$-dependent phase factor because $\hat{U_0}H\hat{U_0}^{-1}$ would still have the $y$-dependent terms $\Omega_2 e^{-2 i K y}$ and $\Omega_2 e^{2 i K y}$.

\section{\label{appendix::PhasesAxialFlux}Phases in Hamiltonians and synthetic axial magnetic flux}
Now, we take into account phases associated with the Raman, microwave, and RF couplings, denoting them respectively as {$\theta_\text{R}$}, $\theta_{1,2}$, and $\theta_\text{RF}$.
Without loss of generality, we consider Hamiltonians without detunings, $\deltaR=\delta_1=\delta_2=0$.

\subsection{Hall cylinder with the nonsymmorphic symmetry}
We refer the reader to Fig.~\ref{Fig1}(b). In this case, we show that $\theta_\text{R}$ and $\theta_{1,2}$ can be gauged away and thus have no effect on the band structure.  
The Hamiltonian $H$ in Eq.~(\ref{H_static_realspace}) becomes
\begin{equation}
H=\frac{\hat{p}_{y}^{2}}{2 m} \mathrm{I}
+\begin{pmatrix}
0 & \frac{\Omega_{\mathrm{R} 2}}{2} e^{-i K y} e^{i \theta_\text{R}} & 0 & \frac{\Omega_{2}}{2} e^{-i \theta_{2}} \\
\frac{\Omega_{\mathrm{R} 2}}{2} e^{i K y} e^{-i
\theta_\text{R}}
&\varepsilon_{0} & \frac{\Omega_{1}}{2} e^{-i \theta_{1}} & 0 \\
0 & \frac{\Omega_{1}}{2} e^{i \theta_{1}} & \varepsilon_{0} & \frac{\Omega_{\mathrm{R} 1}}{2} e^{-i K y} e^{i \theta_\text{R}} \\
\frac{\Omega_{2}}{2} e^{i \theta_{2}} & 0 & \frac{\Omega_{\mathrm{R}
1}}{2} e^{i K y} e^{-i \theta_\text{R}}&0
\end{pmatrix}
\end{equation}
We apply a unitary transformation to $H$:
\begin{equation}
\hat{U}_1^{-1}H\hat{U}_1=\frac{\hat{p}_{y}^{2}}{2 m} \mathrm{I}
+\left(\begin{array}{cccc}
0 & \frac{\Omega_{\mathrm{R} 2}}{2} e^{-i K y} e^{i \Delta \theta_\text{t}} & 0 & \frac{\Omega_{2}}{2}  \\
\frac{\Omega_{\mathrm{R} 2}}{2} e^{i K y} e^{-i \Delta
\theta_\text{t}}
&\varepsilon_{0} & \frac{\Omega_{1}}{2}  & 0 \\
0 & \frac{\Omega_{1}}{2}  & \varepsilon_{0} & \frac{\Omega_{\mathrm{R}
1}}{2} e^{-i K y} e^{i \Delta \theta_\text{t}} \\
\frac{\Omega_{2}}{2}  & 0 & \frac{\Omega_{\mathrm{R}
1}}{2} e^{i K y} e^{-i \Delta \theta_\text{t}}&0\end{array}\right).
  \label{<+label+>}    
\end{equation}
Here \begin{equation}
\label{<+label+>}
\hat{U}_1=
\begin{pmatrix}
    e^{i (\theta_{1}/2-\theta_2/2 )}&0&0&0\\
  0&1&0&0\\
    0&0&e^{i \theta_{1}} &0\\
    0&0&0&e^{i (\theta_{1}/2+\theta_2/2 )}
\end{pmatrix}
\end{equation}
and $\Delta\theta_\text{t}=\theta_\text{R}-\theta_1/2+\theta_2/2$.
Let $y'=y-\Delta\theta_\text{t}/K$. 
The final Hamiltonian is then independent of all the phases {and equivalent to Eq.~({\ref{H_static_realspace})}}.
Note that $\Delta\theta_\text{t}$ is simply half of the accumulated phase acquired by an atom completing a close trajectory counterclockwise in the synthetic dimension [Fig.~\ref{Fig1}(b)], corresponding to a synthetic axial magnetic flux
$\Phi_\text{axial}/\Phi_0=\Delta\theta_\text{t}/\pi$.
In conclusion, $\Phi_\text{axial}$ ($\Delta\theta_\text{t}$) here can only lead to a spatial translation in $y$ and has no effect on the band structure.

\subsection{Hall cylinder with a broken nonsymmorphic symmetry}
We refer the reader to Fig.~\ref{Fig3}(a). In this case, we show that $\theta_\text{R}$ only causes a spatial translation in $y$ and can still be gauged away, 
while an axial phase $\theta_\text{axial}=2\theta_{\mathrm{RF}}+\theta_1-\theta_{2}$ cannot be gauged away and thus can affect the band structure.
The Hamiltonian $H'$ in Eq.~(\ref{H'}) becomes
\begin{align}
&H'=\frac{\hat{p}_{y}^{2}}{2 m} \mathrm{I}\nonumber\\
&+\begin{pmatrix}
0 & \frac{\Omega_{\mathrm{R} 2}}{2} e^{-i K y} e^{i \theta_\text{R}}+\frac{\Omega_{\mathrm{RF2} }}{2} e^{-i \theta_{\mathrm{RF} }} & 0 & \frac{\Omega_{2}}{2} e^{-i \theta_{2}} \\
\frac{\Omega_{\mathrm{R} 2}}{2} e^{i K y} e^{-i\theta_\text{R}}+\frac{\Omega_{\mathrm{RF2} }}{2} e^{i \theta_{\mathrm{RF} }} & \varepsilon_{0} & \frac{\Omega_{1}}{2} e^{-i \theta_{1}} & 
0 \\
0 & \frac{\Omega_{1}}{2} e^{i \theta_{1}} & \varepsilon_{0} & 
\frac{\Omega_{\mathrm{R} 1}}{2} e^{-i K y} e^{i \theta_\text{R}}+\frac{\Omega_{\mathrm{RF1} }}{2} e^{-i \theta_{\mathrm{RF} }} \\
\frac{\Omega_{2}}{2} e^{i \theta_{2}} & 0 & \frac{\Omega_{\mathrm{R} 1}}{2} e^{i K y} e^{-i \theta_\text{R}}+\frac{\Omega_{\mathrm{RF1} }}{2} e^{i \theta_{\mathrm{RF} }} & 0    
\end{pmatrix}
\end{align}
Let $y'=y-\theta_\text{R}/K+\theta_{\mathrm{RF}}/K$, 
the Hamiltonian $H'$ becomes
\begin{equation}
H'=\frac{\hat{p}_{y}^{2}}{2 m} \mathrm{I}+
\left(\begin{array}{cccc}
0 & (\frac{\Omega_{\mathrm{R} 2}}{2} e^{-i K y} +\frac{\Omega_{\mathrm{RF2}
}}{2}) e^{-i \theta_{\mathrm{RF} }} & 0 & \frac{\Omega_{2}}{2} e^{-i \theta_{2}} \\
(\frac{\Omega_{\mathrm{R} 2}}{2} e^{i K y} +\frac{\Omega_{\mathrm{RF2} }}{2})
e^{i \theta_{\mathrm{RF} }} &\varepsilon_{0} & \frac{\Omega_{1}}{2} e^{-i \theta_{1}} & 0 \\
0 & \frac{\Omega_{1}}{2} e^{i \theta_{1}} & \varepsilon_{0} &
(\frac{\Omega_{\mathrm{R} 1}}{2} e^{-i K y} +\frac{\Omega_{\mathrm{RF1} }}{2}) e^{-i \theta_{\mathrm{RF} }} \\
\frac{\Omega_{2}}{2} e^{i \theta_{2}} & 0 & (\frac{\Omega_{\mathrm{R}1}}{2} e^{i K y} +\frac{\Omega_{\mathrm{RF1} }}{2}) e^{i\theta_{\mathrm{RF}}} &0
\end{array}\right),
\end{equation}
which does not depend on $\theta_\text{R}$.
Apply a unitary transformation $\hat{U}_2$ to $H'$:
\begin{equation}
\hat{U}_2^{-1} H'\hat{U}_2
=\frac{\hat{p}_{y}^{2}}{2 m} \mathrm{I}
+\left(\begin{array}{cccc}
0 & (\frac{\Omega_{\mathrm{R} 2}}{2} e^{-i K y} +\frac{\Omega_{\mathrm{RF2}
}}{2})  & 0 & \frac{\Omega_{2}}{2} e^{i \theta_\text{axial}} \\
(\frac{\Omega_{\mathrm{R} 2}}{2} e^{i K y} +\frac{\Omega_{\mathrm{RF2} }}{2})
 &\varepsilon_{0} & \frac{\Omega_{1}}{2}  & 0 \\
0 & \frac{\Omega_{1}}{2}  & \varepsilon_{0} &
(\frac{\Omega_{\mathrm{R} 1}}{2} e^{-i K y} +\frac{\Omega_{\mathrm{RF1} }}{2})  \\
\frac{\Omega_{2}}{2} e^{-i \theta_\text{axial}} & 0 & (\frac{\Omega_{\mathrm{R}
1}}{2} e^{i K y} +\frac{\Omega_{\mathrm{RF1} }}{2}) &0
\end{array}\right),
  \label{eq:axialphaseMatrix}
\end{equation}
where $\theta_\text{axial}=2\theta_{\mathrm{RF}}
+\theta_1-\theta_{2}$ and 
\begin{equation}
\hat{U}_2=
\left(  \begin{array}{cccc}
    1&0&0&0\\
    0&e^{i \theta_{\mathrm{RF} }} &0&0\\
    0&0&e^{i (\theta_{\mathrm{RF}} +\theta_1)} &0\\
    0&0&0&e^{i (2\theta_{\mathrm{RF}} +\theta_1)}\\
\end{array}\right).
  \label{<+label+>}
\end{equation}
The transformed Hamiltonian depends on a single phase, $\theta_\text{axial}$.
Thus, here the synthetic axial magnetic flux $\Phi_\text{axial}/\Phi_0=\theta_\text{axial}/2\pi$
is crucial as it can affect the band structure.

\vspace{5mm}
\twocolumngrid
\section{\label{appendix::radialflux}Importance of the synthetic radial magnetic flux}
As mentioned in the main text, the net radial magnetic flux is key to many phenomena emerging on the Hall cylinder, which otherwise disappear.
To understand this, for example, one can realize a periodic boundary condition by replacing the Raman couplings in Fig.~\ref{Fig1}(b) with RF couplings, which do not change the momentum of an atom. 
Such a cyclic coupling delivers a cylinder without any magnetic field on the cylindrical surface. 
The corresponding Hamiltonian is similar to $H$ but without the $y$-dependent phase factors, i.e. $e^{\pm iKy}=1$. 
{The corresponding dispersion remains parabolic and non-periodic.
Consequently, there are no Bloch oscillations and those observed phenomena in the main text vanish.}
 
Interestingly, one may realize another periodic boundary condition by applying two different pairs of Raman lasers in Fig.~\ref{Fig1}(b)
such that the matrix element $\langle3|H|4\rangle$ ($\langle4|H|3\rangle$) in Eq.~(1) 
changes from $e^{-iKy}$ ($e^{iKy}$)to $e^{iKy}$ ($e^{-iKy}$).
Consequently, a cylinder is penetrated by magnetic fields, but the net radial magnetic flux is zero. 
In this case, the $y$-dependent phase factors in the corresponding Hamiltonian can be gauged away, and the observed phenomena {in the main text} disappear. 
This again uncovers the essence of the net radial magnetic flux for the emergence of the observed phenomena.

\section{\label{appendix::symmetries_H}Symmetries of the Hamiltonian \boldmath{$H$}}
\subsection{Generalized inversion symmetry and band symmetry}
When $\deltaR=\delta_1=\delta_2=0$ and $\Omega_\text{R1}=\Omega_\text{R2}$, the Hamiltonian $H$ in Eq.~(\ref{H_static_realspace}) is invariant under a generalized inversion symmetry, i.e., a spatial inversion ($y\to -y$) followed by a spin inversion ($\ket{1},\ket{2},\ket{3},\ket{4}\to \ket{4},\ket{3},\ket{2},\ket{1}$).
This generalized inversion symmetry guaranties that the band structure or energy spectrum $E(q_y)$ is symmetric with respect to $q_y$, i.e., 
\begin{equation}
E(q_y)=E(-q_y).
  \label{invene}
\end{equation}
However, in general, $\Omega_\text{R1}\neq\Omega_\text{R2}$, which gives rise to 
asymmetric band structures with respect to $q_y$ as verified by numerical calculations.
We also provide a mathematical argument below.

The generalized inversion symmetry operator is written as $\hat{I}=\hat{I_s}\hat{I_y}$,
where $\hat{I_y}$ is the spatial inversion operator that replaces $y$ by $-y$ and $\hat{I_s}$
replaces spin $4,3,2,1$ by spin $1,2,3,4$.
Considering the Hamiltonian $H$ in Eq.~(\ref{H_static_realspace}) with $\deltaR=\delta_1=\delta_2=0$,
we define
\begin{eqnarray}
  &&H_0=
\frac{\hat{p}_{y}^{2}}{2 m} \mathrm{I}\nonumber\\
&&+\left(\begin{array}{cccc}
    0 & \frac{\bar{\Omega}}{2} e^{-i K y}  & 0 & \frac{\Omega_{2}}{2}  \\
  \frac{\bar{\Omega}}{2} e^{i K y} 
&\varepsilon_{0} & \frac{\Omega_{1}}{2}  & 0 \\
0 & \frac{\Omega_{1}}{2}  & \varepsilon_{0} & \frac{\bar{\Omega}}{2} e^{-i K y}  \\
\frac{\Omega_{2}}{2}  & 0 & \frac{\bar{\Omega}}{2} e^{i K y} &0
\end{array}\right)
\end{eqnarray}
and $H_0'=H-H_0$
\begin{equation}
=\left(\begin{array}{cccc}
    0 & \frac{\delta{\Omega}}{2} e^{-i K y}  & 0 & 0 \\
  \frac{\delta{\Omega}}{2} e^{i K y} 
&0 & 0 & 0 \\
0 & 0  & 0 & -\frac{\delta{\Omega}}{2} e^{-i K y}  \\
0  & 0 & -\frac{\delta{\Omega}}{2} e^{i K y} &0
\end{array}\right)
\end{equation}
where $\bar{\Omega}=(\Omega_\text{R1}+\Omega_\text{R2})/2$ and $\delta{\Omega}=(\Omega_\text{R1}-\Omega_\text{R2})/2$.
We readily see that $H_0=\hat{I}^{-1}H_0\hat{I}$ 
and $H_0'=-\hat{I}^{-1}H_0'\hat{I}$.
In momentum space, 
$\hat{I}$ replaces $K$ by $-K$ and spin $4,3,2,1$ by spin $1,2,3,4$,
and we obtain $H_0(-q_y)=\hat{I}^{-1}H_0(q_y)\hat{I}$ 
and $H_0'(-q_y)=-\hat{I}^{-1}H_0'(q_y)\hat{I}$.

Thus, if $\psi(q_y)$ is an eigenstate of $H_0(q_y)$ with an eigenvalue $E_0$, i.e., $H_0(q_y)\psi(q_y)= E_0(q_y) \psi(q_y)$,
then $H_0(-q_y) \hat{I}^{-1}\psi(q_y)=E_0(q_y) \hat{I}^{-1} \psi(q_y)$.
This means that $\hat{I}^{-1} \psi(q_y)$ is an eigenstate of $H_0(-q_y)$ with the same eigenvalue $E_0$.
This shows that the band structure of $H_0$ is inversion symmetric with respect to $q_y=0$.

Recall that $H = H_0 + H_0'$.
The band structure of $H$ is simply {that of} $H_0(q_y)$ plus $H_0'(q_y)$.
For a small but finite $\delta\Omega$, $H_0'$ can be treated as a perturbation, 
and its contribution to $H$ can be estimated by the first-order perturbation theory.
At $q_y$, such an energy contribution to $H$ is 
$\delta E_0(q_y)=\braket{\psi(q_y)|H_0'(q_y)|\psi(q_y)}$, while at $-q_y$
it is opposite, i.e., 
$\delta E_0(-q_y)=\braket{\hat{I}^{-1}\psi(q_y)|H_0'(q_y)|\hat{I}^{-1}\psi(q_y)}=-\delta E_0(q_y)$.
Therefore, the band structure of $H$ is not inversion symmetric even though
$H_0(q_y)$ is inversion symmetric, 
unless when $\Omega_\text{R1}=\Omega_\text{R2}$ ($\delta{\Omega}=0$)
is fulfilled such that $H_0'$ vanishes.

\subsection{Nonsymmorphic symmetry and band crossings}
The Hamiltonian $H$ {in Eq.~({\ref{H_static_realspace}}) as well as Eq.~(1)} is also invariant under a nonsymmorphic symmetry, which comprises a translational operation $\hat{G}=e^{-i
  \hat{p} 
d/(2\hbar)}$ ($d=2\pi/\lambda$) followed by a unitary transformation $\hat{U}$ given by
\begin{equation}
  \hat{U}=\left(
\begin{array}{cccc}
 1 & 0 & 0 & 0 \\
 0 & -1 & 0 & 0 \\
 0 & 0 & -1 & 0 \\
 0 & 0 & 0 & 1 \\
\end{array}
\right).
  \label{transu}
\end{equation}
That is, $\hat{U}\hat{G}H\hat{G}^{-1}\hat{U}^{-1}=H$.
Defining the nonsymmorphic symmetry operator $\hat{S}=\hat{U}\hat{G}$, we readily obtain $[\hat{S},H]=0$, which implies that $\hat{S}$ and $H$ share the same set of eigenstates. The physical meanings of $\hat{U}$ and $\hat{G}$ are explained below. 
First, the translational operator $\hat{G}$ can be understood as shifting the entire $y$ coordinate
to $y+d/2$ by half the period ($d$) of $H$.
Applying $\hat{G}$ to $H$, {i.e., $\hat{G}H\hat{G}^{-1}$,} the matrix elements $\braket{1|H|2}$, $\braket{2|H|1}$, $\braket{3|H|4}$, and $\braket{4|H|3}$
flip their sign. 
Second, the unitary transformation $\hat{U}$ can be understood as flipping the sign
of the second and third spin states. 
Applying $\hat{U}$ to {$\hat{G}H\hat{G}^{-1}$, i.e., $\hat{U}\hat{G}H\hat{G}^{-1}\hat{U}^{-1}$}, the matrix elements {$\braket{1|\hat{G}H\hat{G}^{-1}|2}$, $\braket{2|\hat{G}H\hat{G}^{-1}|1}$, $\braket{3|\hat{G}H\hat{G}^{-1}|4}$, and $\braket{4|\hat{G}H\hat{G}^{-1}|3}$}
flip their sign. The Hamiltonian after these two symmetry
operations ($\hat{U}$ and $\hat{G}$) thus returns to the original Hamiltonian $H$. 

{$\hat{S}^2(=\hat{G}^2)$} is a translational operator corresponding to a shift of $d$ in the $y$ coordinate, such that $[\hat{S}^2, H] =0$. Therefore, the Hamiltonian $H$ is invariant after a shift of $d$ in $y$, a discrete translational symmetry. The eigenvalues of $H$ thus have a periodicity of $d$ in $y$. 
The eigenwavefunctions of $H$ can be written in the form of Bloch waves, $e^{i  q_y y} w(y)$, where $w(y)$ has a period of $d$.
Since the nonsymmorphic symmetry operator $\hat{S}$ and the Hamiltonian $H$ share the same set of eigenstates, 
we can construct the eigenstates of $H$ (and $\hat{S}$) in the
following two types (in the form of Bloch waves) by considering the physical meanings of $\hat{S}$ mentioned above:
\begin{eqnarray}
  \psi_p(q_y)=e^{i q_y y}(
  &&u_1(y)\ket{1}+u_2(y)\ket{2}\nonumber\\
  +&&u_3(y)\ket{3}+u_4(y)\ket{4}),   
  \label{eigenp}
\end{eqnarray} 
\begin{eqnarray}
  \psi_m(q_y)=e^{i q_y y} (&&v_1(y)\ket{1}+v_2(y)\ket{2}\nonumber\\
  +&&v_3(y)\ket{3}+v_4(y)\ket{4}).
  \label{eigenm}
\end{eqnarray}
Here, 
\begin{align}
\label{eigensol1}
u_1(y+d/2)-u_1(y)&=u_2(y+d/2)+u_2(y)&&\nonumber\\
=u_3(y+d/2)+u_3(y)&=u_4(y+d/2)-u_4(y)=0,
\end{align}
\begin{align}
\label{eigensol2}
v_1(y+d/2)+v_1(y)&=v_2(y+d/2)-v_2(y)&&\nonumber\\
=v_3(y+d/2)-v_3(y)&=v_4(y+d/2)+v_4(y)=0.
\end{align}
Applying $\hat{S}$ to Eqs.~(\ref{eigenp}, \ref{eigenm}), one can verify that $\psi_p$ and $\psi_m$ are eigenfunctions of $\hat{S}$ with the corresponding eigenvalues $\pm e^{i q_y d/2}$.
With Eqs.~(\ref{eigensol1}, \ref{eigensol2}), we also see that $\psi_p(q_y)$ and $\psi_m(q_y)$ are still Bloch waves labeled by $q_y$.

Consider two sets of eigenfunctions $\{\psi_p(q_y),\psi_m(q_y)\}$ and $\{\psi_p(q_y+K),\psi_m(q_y+K)\}$. 
Their corresponding eigenvalues of the operator $\hat{S}$ are $\{e^{i q_y d/2},-e^{i q_y d/2}\}$ and $\{-e^{i q_y d/2},e^{i q_y d/2}\}$.
Thus, one obtains $\psi_p(q_y)=\psi_m(q_y+K)$ and $\psi_m(q_y)=\psi_p(q_y+K)$.
This suggests two properties associated with the nonsymmorphic symmetry: (I) both $\psi_p(q_y)$ and $\psi_m(q_y)$ have a periodicity of
$2K$ in $q_y$, and (II) $\psi_p(q_y)$ and $\psi_m(q_y)$ are offset from each other by $K$ in $q_y$. 
Denote the corresponding eigenenergies (eigenvalues of $H$) for $\psi_p(q_y)$ and $\psi_m(q_y)$ as $E_p$ and $E_m$, the energy spectrum also possesses properties associated with the nonsymmorphic symmetry, corresponding to properties (I) and (II) above. Corresponding to (I), we have
\begin{equation}
  E_p(q_y)=E_m(q_y+K).
  \label{sene}
\end{equation}
This suggests that the band structure has crossing points at some $q_y$.
Recall that the Hamiltonian $H$ with $\deltaR=\delta_1=\delta_2=0$ and $\Omega_\text{R1}=\Omega_\text{R2}$ possesses a generalized inversion symmetry in Eq.~(\ref{invene}).
Given the relations in Eq.~(\ref{invene}) and Eq.~(\ref{sene}), we obtain
\begin{equation}
  E_p(q_y)=E_m(-q_y+K).
  \label{<+label+>}
\end{equation}
Consequently, for $q_y=(2n+1)K/2$ where $n$ is an integer, $E_p$ is equal to $E_m$, corresponding to a degenerate point (band crossing) in the band structure.
Such a degeneracy at $q_y=(2n+1)K/2$ is protected by the nonsymmorphic symmetry and the generalized inversion symmetry. 
If any of $\deltaR, \delta_1, \delta_2$ is nonzero or 
if $\Omega_\text{R1}\neq\Omega_\text{R2}$, the generalized inversion symmetry is broken while the nonsymmorphic symmetry is retained, the two branches still cross but at $q_y\neq (2n+1)K/2$.

Furthermore, the two independent branches in the spin-mechanical momentum coupling scheme in Fig.~\ref{Fig2}(a) implies that the plane wave basis $\{\ket{\hbar(q_y+nK);m}\}$ in Eq.~(\ref{plane_wave_basis}) can also be decomposed into two subsets based on the nonsymmorphic symmetry. 
These two branches can be written in the following form:
\begin{align}
\label{phi1}
&\phi_p(q_y)=\sum_n(c_{1,n}\ket{q_y +2nK;1}+c_{2,n}\ket{q_y +2nK+K;2}\nonumber\\
+&c_{3,n}\ket{q_y +2nK+K;3}+c_{4,n}\ket{q_y +2nK;4})
\end{align}
and
\begin{flalign}
\label{phi2}
&\phi_m(q_y)=\sum_n(d_{1,n}\ket{q_y + 2nK+K;1}+d_{2,n}\ket{q_y +2nK;2}\nonumber\\
+&d_{3,n}\ket{q_y + 2nK;3}+d_{4,n}\ket{q_y + 2nK+K;4}).
\end{flalign}
Equating Eqs.~(\ref{phi1}, \ref{phi2}) with Eqs.~(\ref{eigenp}, \ref{eigenm}) respectively, the coefficients in the above equations satisfy
\begin{align}
  &\sum_n c_{1,n}e^{i 2nK y}=u_1(y),\qquad
  \sum_n c_{2,n}e^{i (2nK+K) y}=u_2(y),&&\nonumber\\
  &\sum_n c_{3,n}e^{i (2nK+K) y}=u_3(y),\
  \sum_n c_{4,n}e^{i 2nK y}=u_4(y),
\end{align}
and
\begin{align}
  &\sum_n d_{1,n}e^{i (2nK+K) y}=v_1(y),\
  \sum_n d_{2,n}e^{i 2nK y}=v_2(y),&&\nonumber\\
  &\sum_n d_{3,n}e^{i 2nK y}=v_3(y),\qquad
  \sum_n d_{4,n}e^{i (2nK+K) y}=v_4(y).
\end{align}
From Eqs.~(\ref{phi1}, \ref{phi2}), we readily see that $\phi_p(q_y)$ and $\phi_m(q_y+K)$ are identical if one equates
$d_{1,n}$ with $c_{1,n+1}$, 
$d_{4,n}$ with $c_{4,n+1}$, 
$d_{2,n}$ with $c_{2,n}$, 
and 
$d_{3,n}$ with $c_{3,n}$.
Thus, Eqs.~(\ref{phi1}, \ref{phi2}) respectively correspond to the band 1 and band 2 in Fig.~\ref{Fig2} in the main text, providing another way to understand band crossings due to the nonsymmorphic symmetry.

\section{\label{appendix::symmetries_Hprime}Symmetries of the Hamiltonian \boldmath{$H'$}}
When the RF wave is applied, 
the corresponding Hamiltonian $H'$ [Eq.~{(\ref{H'})}] is obtained by adding the RF couplings to the Hamiltonian $H$. These $y$-independent RF terms are added to the Raman terms in the $\langle 1|H|2\rangle$, $\langle 2|H|1\rangle$, $\langle 3|H|4\rangle$, $\langle 4|H|3\rangle$ matrix elements of $H$. Upon $d/2$ translation by $\hat{G}$, the Raman terms flip sign but the $y$-independent RF terms do not; therefore, the nonsymmorphic symmetry is broken in $H'$.  
However, a $d$ translation still leaves $H'$ invariant.
Similar to $H$, the generalized inversion symmetry is also broken in $H'$ and thus the band structure is asymmetric.

\section{\label{appendix::realspace_wavefunction}Calculations of BEC wavefunctions in the real space}
We solve the eight-spin version (Appendix \ref{appendix::H_8spins}) of the Hamiltonians $H_{q_y}$ or $H'_{q_y}$ [see Eqs.~(\ref{H_Momentum}) and (\ref{An_RF})]
to obtain the probability amplitude ($b_{n,m}^{q_y}$) of the constituent plane waves of the form $b_{n,m}^{q_y}e^{i(q_y+ nK)y}\ket{m}$, whose superposition gives the BEC wavefunction in the real space. 
From the wavefunction, we obtain the variations of the density and phase in the real space for each spin state.
For example, we perform such calculations for a BEC at $q_y=0$ in the ground or first excited bands in two exemplary cases (in units of $\E_r$).
(1) A band structure with band crossings at $\Omega_\text{R1}=2.3$, $\Omega_1=2.3$,
and other Raman and microwave couplings obtained by the corresponding scaling relations. No RF couplings.
(2) A band structure with gap opening at $\Omega_\text{R1}=2.3$, 
$\Omega_1=2.3$, $\Omega_\text{RF1}=0.8$, $\theta_\text{axial}=0$,
and other Raman, microwave, and RF couplings obtained by the corresponding scaling relations (Appendix \ref{appendix::H_8spins}). The results shown below focus on spin states $\ket{1}$, $\ket{2}$, $\ket{3}$, and $\ket{4}$.

\begin{figure}[ht!]
\includegraphics[width=3.2in]{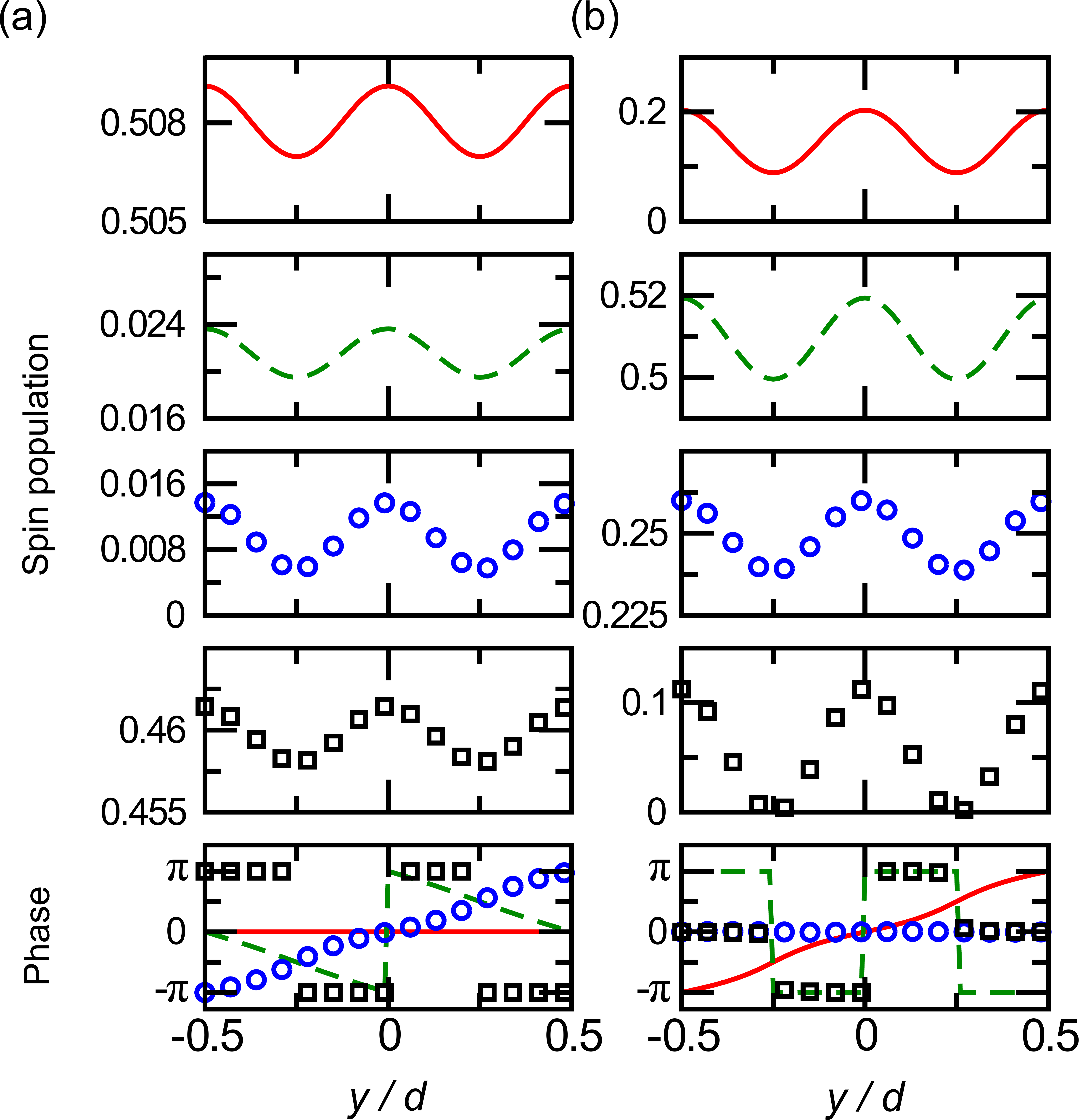}
\caption{\small{
Spin population and phase versus $y$ for case 1 (see the text).
Calculations are performed for a BEC at $q_y=0$ in the (a) ground band and 
(b) first excited band.
The red line, green dashed line, blue circles, and black squares correspond to
the spin states $\ket{1}$, $\ket{2}$, $\ket{3}$, and $\ket{4}$, respectively.
The plotted population of the spin component $i$, $\rho_i$, is normalized by the condition $\sum_i\int_0^{1}\rho_i\text{d}(y/d)=1$ (also used for Fig.~{\ref{FigSIBECwavefunction_regular}}).
}
}
\label{FigSIBECwavefunction_Mobius}
\end{figure}

\vspace{3mm}
\begin{figure}[ht!]
\includegraphics[width=3.2in]{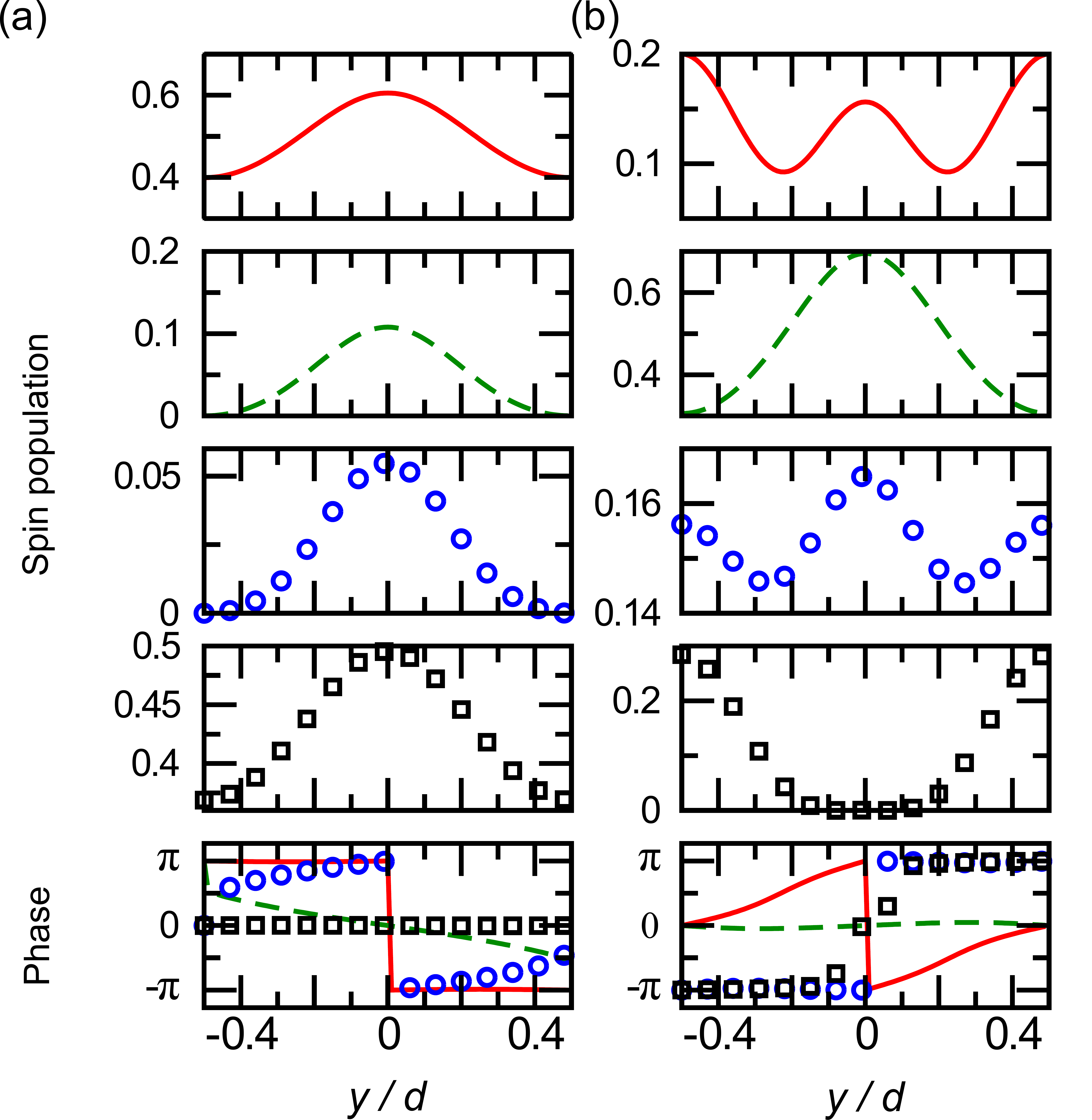}
\caption{\small{
Spin population and phase versus $y$ for case 2 (see the text).
Calculations are performed for a BEC at $q_y=0$ in the (a) ground band and 
(b) first excited band.
The red line, green dashed line, blue circles, and black squares correspond to
$\ket{1}$, $\ket{2}$, $\ket{3}$, and $\ket{4}$, respectively.
}
}
\label{FigSIBECwavefunction_regular}
\end{figure}

\vspace{3mm}
\noindent
(1) \textit{Case 1}\\
The calculated population and phase in the real space for each spin state shown in Fig.~\ref{FigSIBECwavefunction_Mobius}(a) and \ref{FigSIBECwavefunction_Mobius}(b) are respectively for a BEC in the ground and the first excited bands. The red line, green dashed line, blue circles and black squares respectively correspond to $\ket{1}$, $\ket{2}$, $\ket{3}$, and $\ket{4}$.
The BEC wavefunction corresponding to Fig.~\ref{FigSIBECwavefunction_Mobius}(a)[(b)] can be described by $\phi_p(q_y=0)$ [$\phi_m(q_y=0)$] in Eq.~(\ref{phi1}) [Eq.~(\ref{phi2})], an eigenfunction of the $\hat{S}$ operator with an eigenvalue of $e^{i q_y d/2}=1$ ($-e^{i q_y d/2}=-1$). 
Because of the nonsymmorphic symmetry,
we find that (I) the calculated population of each spin state has a periodicity of $d/2$; (II) for the ground band, the phase of $\ket{1}$ and $\ket{4}$ ($\ket{2}$ and $\ket{3}$) have a period of $d/2$ ($d$). For the first excited band, the phase of 
$\ket{1}$ and $\ket{4}$ ($\ket{2}$ and $\ket{3}$) have a period of $d$ ($d/2$). 
In general, for $q_y\neq 0$, the phase of two spin states would have a periodicity of $d$ while the phase of the other two would have a periodicity of $d/2$. 
This is because a nonzero $q_y$ only introduces an overall phase to the spin states at $q_y=0$.

\vspace{3mm}
\noindent
(2) \textit{Case 2}\\
The calculated population and phase in the real space for each spin shown in Figs.~\ref{FigSIBECwavefunction_regular}(a) and \ref{FigSIBECwavefunction_regular}(b), respectively, correspond to a BEC in the ground and first excited bands. 
The periodicity of the population and phase for each spin is identical to the periodicity of the Hamiltonian, $d$.
The maximum population of the ground state sits at $y=\pm nd $, where $n$ is an integer, because of the $s$-wave nature of the ground state.
On the other hand, for the first excited state, the maximum population of $\ket{1}$ and $\ket{4}$ sits at $y=\pm (2n+1)d/2 $ rather than $y=\pm nd $. 
Besides, there is also local peak population appearing at $y=\pm nd$.

\section{\label{appendix::eigenstate_calculation}Eigenstate calculations}
We use the eight-spin version (Appendix \ref{appendix::H_8spins}) of the Hamiltonians $H_{q_y}$ and $H'_{q_y}$ [see Eqs.~(\ref{H_Momentum}), (\ref{An_sameMomentum}), and (\ref{An_RF})] to calculate the corresponding band structures in Figs.~\ref{Fig2}, \ref{Fig3}, \ref{Fig4} with $n$ ranging from $-13$ to $13$, i.e., each Hamiltonian is a 108-by-108 matrix. 
We use the eight-spin version of $H_{\text{strip}}$ [see Eq.~(\ref{H_unzipped})] to calculate the dispersion in Fig.~\ref{FigSI_Hall_Strip}. We solve the eigenstates of $H_{q_y}$, $H'_{q_y}$, and $H_{\text{strip}}$ as a function of quasimomentum to obtain the corresponding average mechanical momentum and spin compositions. 
These results can be converted to functions of time based on the calibrated relation between quasimomentum and time.
In general, the eigenstate is a normalized vector of the form $(..., b_{n,m}^{q_y}, ...)^T$. 
The coefficient $b_{n,m}^{q_y}$ is the probability amplitude ($|b_{n,m}^{q_y}|^2$ corresponds to the population) of the state $\ket{\hbar(q_y+nK);m}$. 
According to the discussions in Appendix \ref{appendix::H_8spins},
the average mechanical momentum of the eigenstate at $q_y$ 
is determined as $\hbar\sum_{n,m}|b_{n,m}^{q_y}|^2(q_y+nK)/(\sum_{n,m}|b_{n,m}^{q_y}|^2)$, 
where $m=1,2,3,4$. 
The fractional population of spin state $\ket{m}$ at $q_y$ is $\sum_{n}|b_{n,m}^{q_y}|^2/(\sum_{n,m}|b_{n,m}^{q_y}|^2)$, where $m=1,2,3,4$.


\vfill\null
\section{\label{appendix::calibration_quasimom_vs_t}Calibration of quasimomentum\\ versus \boldmath{$\t_hold$}}
\begin{figure}[h]
\includegraphics[width=3.4in]{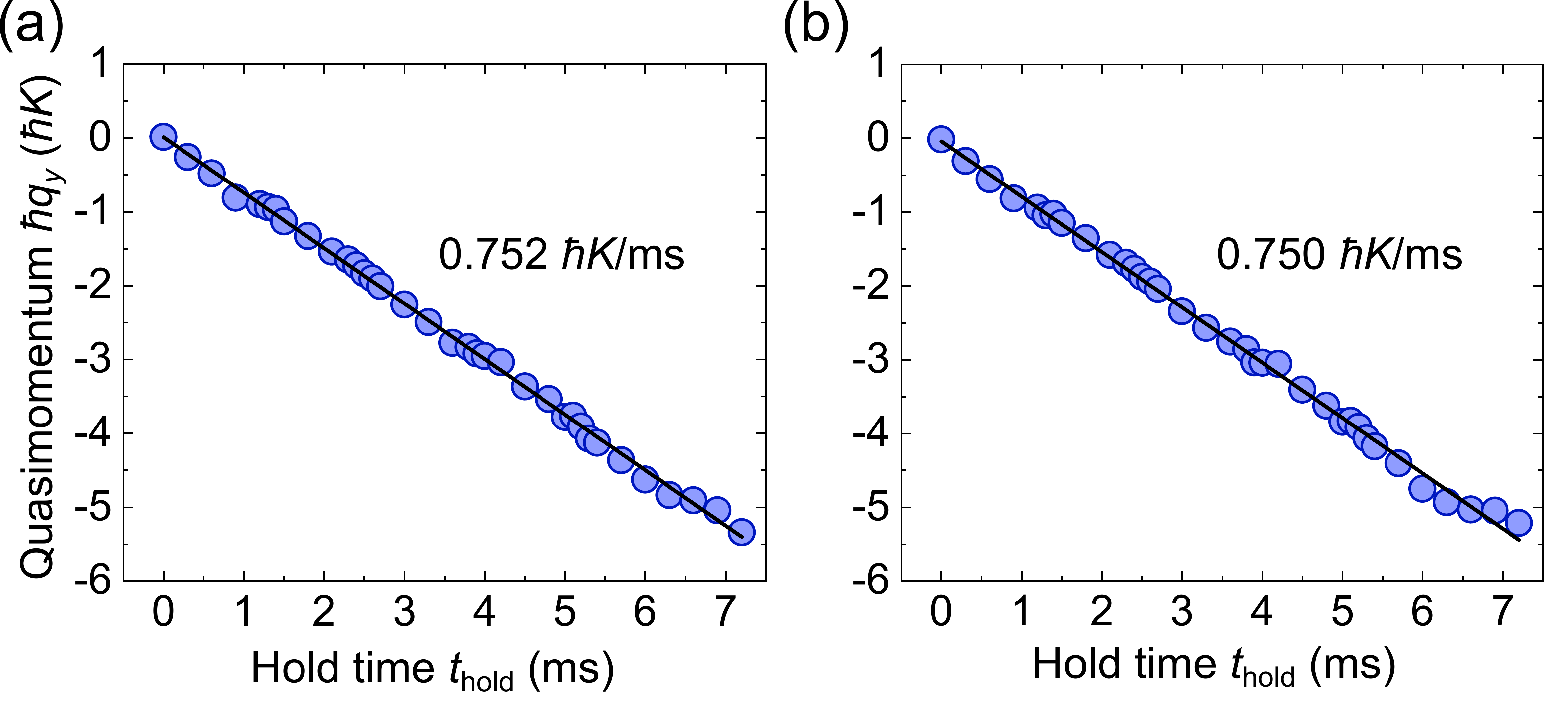}
\caption{\small{
Calibration of quasimomentum versus $\t_hold$. Panels (a) and (b), corresponding to the spin components $\ket{1}$ and $\ket{4}$, respectively, are obtained from the band-1 transport experiment in Fig.~\ref{Fig2}. Dots are experimental data and lines are linear fits.
}
}
\label{FigSI_qyvstime_calibration}
\end{figure}

The quasimomentum of the BEC at $\t_hold$ can be measured by the displacement of the mechanical momentum components of, say $\left|1\right\rangle$ or $\left|4\right\rangle$, at $\t_hold$ relative to those at $\t_hold=0$ ($q_y=0$).
Fig.~\ref{FigSI_qyvstime_calibration} shows the calibrated relation between quasimomentum and $\t_hold$. 
The corresponding slope, $d(\hbar q_y)/d(\t_hold)$, is obtained by the linear fit to the data.
The average slope, $0.751$ $\hbar K/$ms, serves as the calibration used for calculations.
The calculated physical quantities that are functions of quasimomentum can then be converted to functions of $\t_hold$.
Note that the calibrated slope is slightly smaller than the predicted $0.843$ $\hbar K/$ms which is obtained by using $g=9.81$ m/s$^2$ for the gravity.
This small difference may be due to the presence of small background (e.g.~magnetic) fields that counteract the gravity during experiments.


\section{\label{appendix::HallStrip}Transport in a Hall strip}
\begin{figure}[ht!]
\includegraphics[width=3.4in]{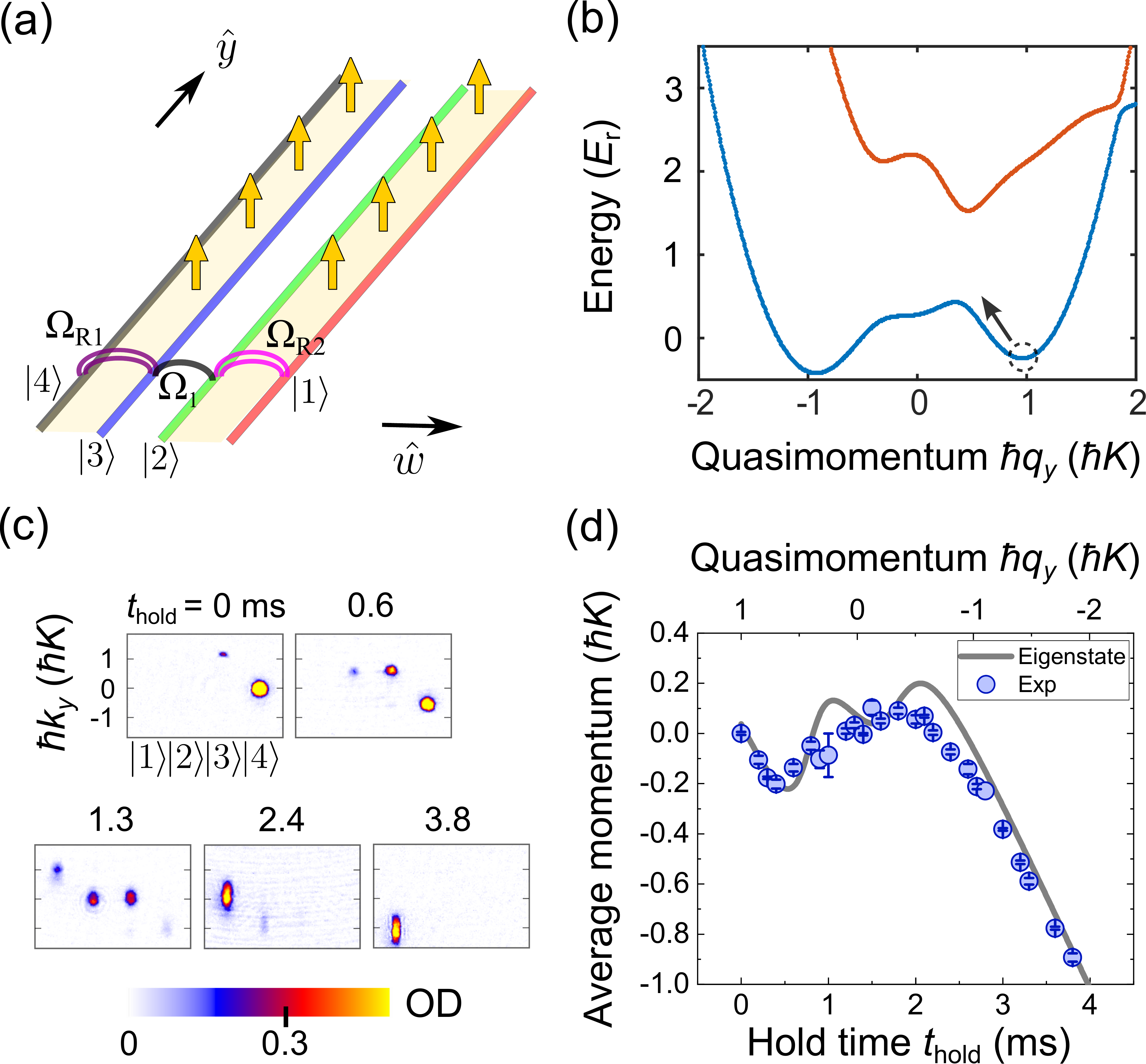}
\caption{\small{
Transport in a Hall strip.
(a) Schematic of a Hall strip. (b) The single-particle dispersion. (c) Select TOF images for the quantum transport measurement with a BEC initially prepared around the minimum of the right well [dashed circle in (b)]. (d) Average momentum versus $\t_hold$ and quasimomentum. Circles are experimental data (error bars are standard errors of typically 5 repetitive measurements). Solid lines are eigenstate calculations.
}
}
\label{FigSI_Hall_Strip}
\end{figure}

To realize a planar Hall strip [Fig.~\ref{FigSI_Hall_Strip}(a)], we remove $\Omega_2$ and keep $\Omega_\text{R1,R2}$ and $\Omega_{1}$ in Figs.~\ref{Fig1}(b) and \ref{Fig1}(c), imposing an open boundary condition along $\hat{w}$.
The strip is pierced by the same magnetic field as for the cylinder.
Nonetheless, as shown in Eqs.~(\ref{eq:HallStripUnitary}) and (\ref{H_unzipped}),
the Raman-imprinted phase factor $e^{\pm iKy}$ can now be gauged away,
resulting in a non-periodic single-particle dispersion [Fig.~\ref{FigSI_Hall_Strip}(b)]. We probe this dispersion by performing the same type of quantum transport measurement with a BEC initially prepared around the minimum of the right well in the ground band [Fig.~\ref{FigSI_Hall_Strip}(b)].

Figure~\ref{FigSI_Hall_Strip}(c) presents select TOF images at various $\t_hold$ and the corresponding quasimomentum. The extracted average momentum is shown in Fig.~\ref{FigSI_Hall_Strip}(d), in which circles are experimental data and solid lines are eigenstate calculations.
In this case, BEC's average momentum keeps increasing due to the gravity.
{No Bloch oscillations or periodic occurrences are observed.} 

\section{\label{appendix::initialstate_Exp}Initial state preparations in experiments}
\noindent
(1) \textit{Band 1 and band 2 in Fig.~\ref{Fig2}(b)}\\
We note that the eigenstate around $q_y=0$ in band 1 (band 2) has dominant populations in $\left|4\right\rangle$ and $\left|1\right\rangle$ ($\left|3\right\rangle$ and $\left|2\right\rangle$). 
Thus, to load a BEC around $q_y=0$ in band 1 (band 2), a BEC is first prepared 
at $\left|4\right\rangle$ ($\left|3\right\rangle$) 
with $\deltaR, \delta_1<-2.5$ $\E_r$. 
The value of $\delta_{2}$ is inferred from Eq.~(\ref{detuning_relation}).

For band 1, we then ramp on the Raman and microwave couplings $\Omega_\text{R1,R2}$ and $\Omega_{1,2}$ from zero to final values while ramping $\deltaR$ and $\delta_1$ to zero in 15 ms. 

For band 2, we then ramp on the microwave couplings $\Omega_{1,2}$ from zero to final values while ramping both $\deltaR$ and $\delta_1$ to around $-0.6$ $\E_r$ in 15 ms. 
Subsequently, while keeping $\Omega_{1,2}$ at the final values, we ramp on the Raman couplings $\Omega_\text{R1,R2}$ from zero to final values in 5 ms during which we ramp $\deltaR$ and $\delta_1$ to zero in 3 ms and then hold $\deltaR$ and $\delta_1$ at zero for the remaining 2 ms.

\vspace{3mm}
\noindent
(2) \textit{Ground band in Figs.~\ref{Fig3}(d) and \ref{Fig4}(a)}\\
A BEC is first prepared at $\left|4\right\rangle$ with $\deltaR, \delta_1<-5$ $\E_r$.
Then, we ramp on the Raman and microwave couplings $\Omega_\text{R1,R2}$ and $\Omega_{1,2}$ from zero to final values while ramping $\deltaR$ and $\delta_1$ to zero in 15 ms. 
At the very beginning (at which $\deltaR, \delta_1<-5$ $\E_r$, so the RF wave is off resonant) of this 15-ms ramp, the RF couplings $\Omega_\text{RF1,RF2}$ are abruptly turned on to the final values. Then, $\Omega_\text{RF1,RF2}$ are held at the same final values while $\deltaR$ and $\delta_1$ are ramped to zero in 15 ms.

\vspace{3mm}
\noindent
(3) \textit{Ground band in Fig.~\ref{FigSI_Hall_Strip}(a)}\\
To prepare a BEC around the minimum of the right well, a BEC is first prepared at $\left|4\right\rangle$ with $\deltaR, \delta_1<-2.5$ $\E_r$. 
Then, we ramp on the Raman and microwave couplings $\Omega_\text{R1,R2}$ and $\Omega_{1}$ from zero to final values while ramping $\deltaR$ and $\delta_1$ to zero in 15 ms. In this case, $\Omega_{2}$ is zero throughout the experiment.

\section{\label{appendix::condensate_frac}Condensate fraction}
In experiments, the condensate fraction before loading atoms to band structures is 90\%, decreases to 50-70\% after a 15-ms loading procedure, and then drops to 40-60\% after another 7-ms holding time for transport. We find that the decreasing of the condensate fraction (with a typical time scale of 40-60 ms) is mainly due to the finite lifetime of atoms in the $F=2$ hyperfine manifold in our system.
\begin{figure*}[t]
\includegraphics[width=6.0in]{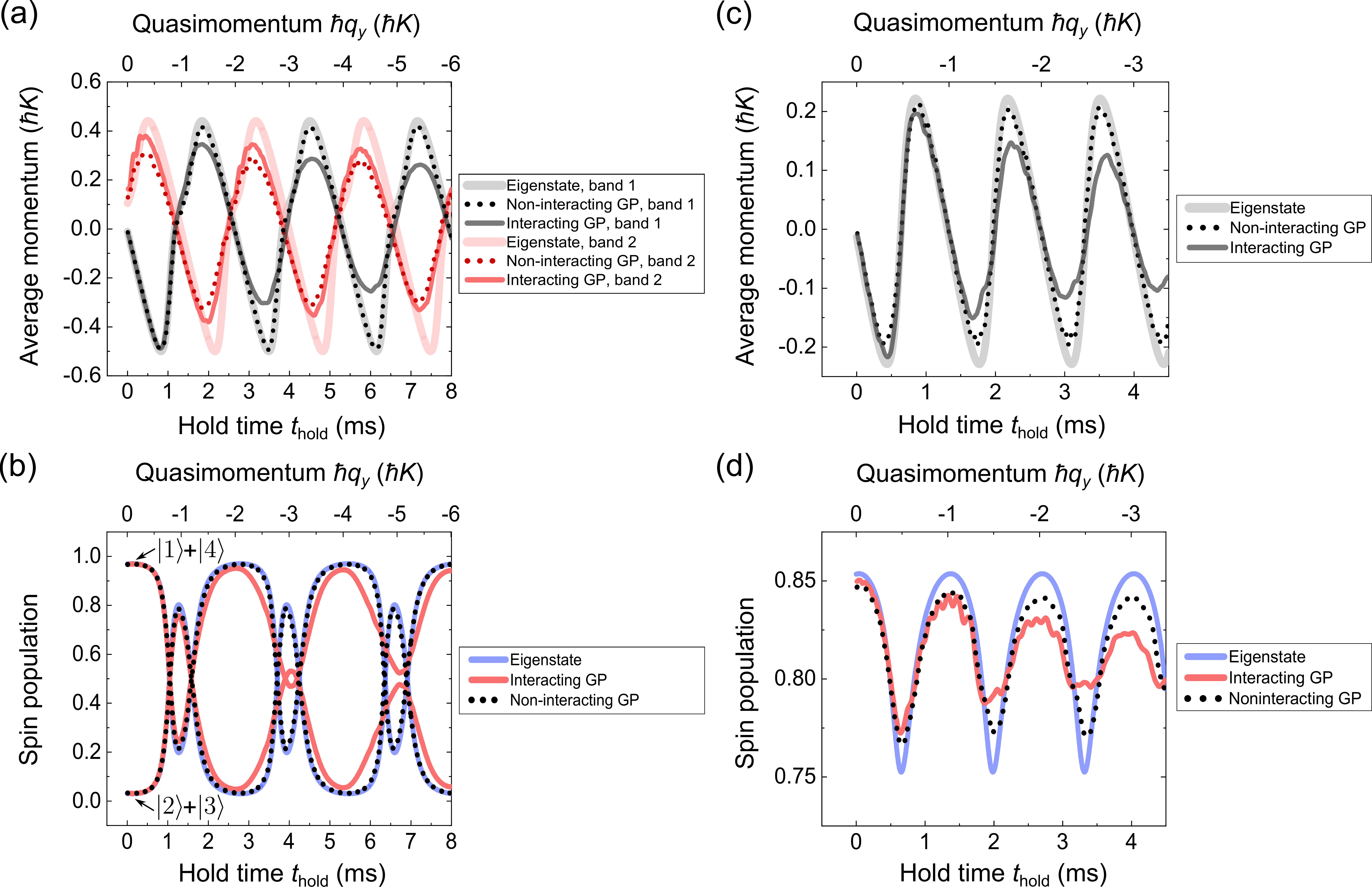}
\caption{\small{
Comparison between GP (interacting and non-interacting) and eigenstate calculations.
(a) Calculations of the average momentum versus $\t_hold$ for the transport in band 1 and band 2 in Fig.~\ref{Fig2}(b). The spin population versus $\t_hold$ corresponding to the transport in band 1 is shown in (b). 
(c) Calculations of the average momentum versus $\t_hold$ for the transport in Fig.~\ref{Fig3}(d). The corresponding spin population versus $\t_hold$ is shown in (d).
}
}
\label{FigSIGPcomparison}
\end{figure*}

\section{\label{appendix::Imaging_analysis}Imaging analysis}
For each spin state in TOF images, a proper window is chosen to enclose the corresponding atomic clouds.
We sum the optical density of each pixel over the entire window to obtain the atom number $N_i$ of spin state $i$.
The average mechanical momentum $p_i$ of atoms in spin state $i$ is determined by the difference in average pixel position between the atoms and a BEC that has zero mechanical momentum. Such a difference is then converted to mechanical momentum based on the calibrated conversion between $\hbar K$ and image pixels.
The average mechanical momentum $p$ of all atoms is the weighted average of the mechanical momentum of each spin state, i.e., $p = N_ip_i/(\sum_{i}N_i)$.

\section{\label{appendix::GP}Gross-Pitaevskii (GP) simulations}
We solve the time-dependent Gross-Pitaevskii equation to simulate BEC's transport with an atom number $N$ of 15000.
We have checked that the small {spin-dependent interaction} has negligible effect in our calculated results, 
so {a spin-independent scattering length} $a_\text{s}=93.2467a_\text{B}$ is used for all the scattering lengths, where $a_\text{B}$ is the Bohr radius.
{The effective Hamiltonian reads
$H_\text{eff}=H+V+U$,
where $H$, $V$, and $U$ respectively correspond to the kinetic energy plus light (Raman, microwave, RF) couplings, harmonic trapping potential, and interactions.}
The interaction term is $U=\frac{4\pi\hbar^2 a_s}{m} n(\vec{r})$, 
where $n(\vec{r})$ is the density of all the spin states and is normalized to $N$, i.e., $\int_r n(\vec{r}) dr=N$.
Recursively applying the Trotter formula,
$e^{ (A+B) \delta t}= e^{A \delta t/2}e^{B \delta t}e^{A \delta t/2} +O(\delta t^3)$,
we obtain an approximation of the short time propagator that is a product of
the exponentials of {$H$, $V$, and $U$}.
The kinetic energy propagator $e^{K \delta t}$ is evaluated in momentum space and the rest in real space.

\vspace{3mm}
\noindent
(1) \textit{Initial state preparation and time evolution}\\
To prepare the initial state of a BEC starting in the ground band in Figs.~\ref{Fig2}, \ref{Fig3}, \ref{Fig4}, we use imaginary time evolution to find the ground state of $H_\text{eff}$. 
That is, an arbitrary initial state $\psi_\text{trial}$ is evolved by
applying the operator $e^{H_\text{eff}\delta t/\hbar}$ until the normalized wave function does not change.
We apply $e^{-i H_\text{eff} \delta t/\hbar}$ consecutively to propagate the wavefunction in real time.

To prepare the initial state of a BEC starting in the first excited band in Fig.~\ref{Fig2},
we follow the same preparation method used in the experiment. 
In this case,
more than 90\% of atoms is loaded to the first excited band.


\vspace{3mm}
\noindent
(2) \textit{Comparison between GP and eigenstate calculations}\\
Figure \ref{FigSIGPcomparison} presents a
comparison between eigenstate and GP (interacting and non-interacting) calculations for the transport in Figs.~\ref{Fig2} and \ref{Fig3}.
Figure \ref{FigSIGPcomparison}(a) compares these calculations of the average momentum versus $\t_hold$ for the transport in band 1 and band 2 in Fig.~\ref{Fig2}(b). 
The spin population versus $\t_hold$ corresponding to the transport in band 1 
is shown in Fig.~\ref{FigSIGPcomparison}(b).
Figure \ref{FigSIGPcomparison}(c) compares these calculations of the average momentum versus $\t_hold$ for the transport in Fig.~\ref{Fig3}(d).
The corresponding spin population versus $\t_hold$ is shown in Fig.~\ref{FigSIGPcomparison}(d).
By comparing the results of interacting and non-interacting GP, we see that interactions lead to damping of both the momentum and spin oscillations. 
In addition, results of non-interacting GP and eigenstate calculations are similar, almost overlapping with each other {except for transport in band 2 in Fig.~{\ref{FigSIGPcomparison}}(a)}.

{For the transport in band 2}, the results of interacting GP are only slightly more damped than that of the non-interacting GP.
Besides, these GP results are notably different from that of the eigenstate calculation. 
This is probably because the ramping procedure used in GP for the initial state preparation in this case is relatively fast {compared to} the time scales of both interactions and {adiabatic loading}. This gives rise to small interaction effects and non-adiabatic loading to band 2.

\section{\label{appendix::calibration_axial_flux}Calibration of the axial phase and axial magnetic flux}
\begin{figure}[h]
\includegraphics[width=2.2in]{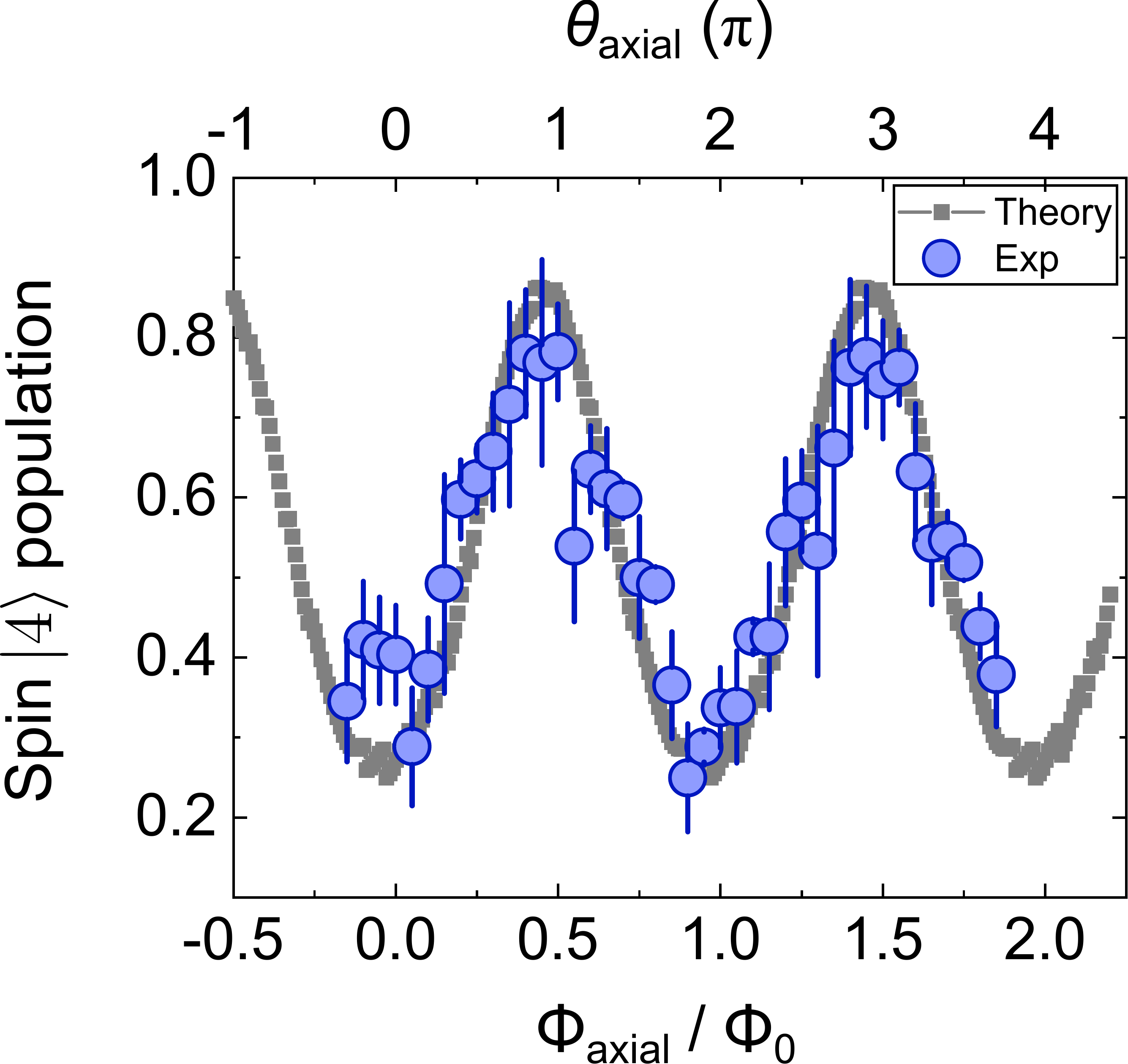}
\caption{\small{
Calibration of the axial phase and axial magnetic flux.
Spin-$\ket{4}$ population versus $\theta_\text{axial}$ ($\Phi_\text{axial}$) at the end of a $90$-$\mu s$ pulse, composed of microwaves and RF waves and applied to a BEC initially prepared at $\ket{4}$.
In this case, $\Omega_{1(2)}=2.2(3.2)$ $\E_r$ and $\Omega_\text{RF1(RF2)}=1.1(-1.6)$ $\E_r$.
Blue circles are experimental data and black dots are numerical results
obtained by solving the time-dependent Schr\"odinger equation. Error bars are standard errors of typically 5 repetitive measurements.
}
}
\label{FigSI_phase_calibration}
\end{figure}
To calibrate the axial phase $\theta_\text{axial}$ and axial magnetic flux $\Phi_\text{axial}$ ($\Phi_\text{axial}/\Phi_0=\theta_\text{axial}/2\pi$) for the experiment in Fig.~\ref{Fig4}, 
we perform quench experiments.
We refer the reader to Fig.~\ref{Fig3}(a). 
We first prepare a BEC at $\ket{4}$, and then suddenly turn on only microwave and RF couplings [$\Omega_{1(2)}=2.2(3.2)$ $\E_r$ and $\Omega_\text{RF1(RF2)}=1.1(-1.6)$ $\E_r$] 
{with corresponding phases $\theta_{1,2}$ and $\theta_\text{RF}$} for $90$ $\mu s$.
This results in spin dynamics that depends on a single phase 
$\theta_\text{axial}=2\theta_{\mathrm{RF}}+\theta_1-\theta_{2}$,
as explained in Eq.~(\ref{eq:axialphaseMatrix}).
The parameters are chosen such that the sensitivity of spin dynamics to $\theta_\text{axial}$ is measurable and allows the phase calibration. 
Specifically, at the end of such a pulse, we measure spin-$\ket{4}$ population as a function of $\theta_\text{axial}$.
The experimental results are then compared with numerical results obtained by solving the time-dependent Schr\"odinger equation to calibrate $\theta_\text{axial}$.
Fig.~\ref{FigSI_phase_calibration} presents a typical calibration, demonstrating our capability of controlling $\theta_\text{axial}$ ($\Phi_\text{axial}$).

To independently control $\theta_{\mathrm{RF}}$ and $\theta_1-\theta_{2}$,
we use an RF function generator with two outputs.
One output channel generates the desired RF field with phase $\theta_{\mathrm{RF}}$ at atoms. 
Another output channel sends an RF signal to a microwave mixer, 
which mixes this RF signal with a microwave (from a microwave generator) 
and then outputs the desired microwave fields 1 and 2, 
where $\theta_1-\theta_{2}$ is controlled by the phase of the RF signal sent to the mixer. 
We thus can control $\theta_\text{axial}$. 
Our RF generator can control the phases of the RF signals within $\pm 1.75$ milliradian. 
For daily experiments, the stability of $\theta_\text{axial}$ is estimated to be within $\pm 0.035 \pi$ (including the fitting uncertainty of $\pm 0.025 \pi$) 
obtained by monitoring the drift of the experimental curve (Fig.~{\ref{FigSI_phase_calibration}}) in a day.


\section{\label{appendix::effects_interactions_LZ}Effects of inter-particle interactions on Landau-Zener tunneling}
\begin{figure}[t!]
\includegraphics[width=3.4in]{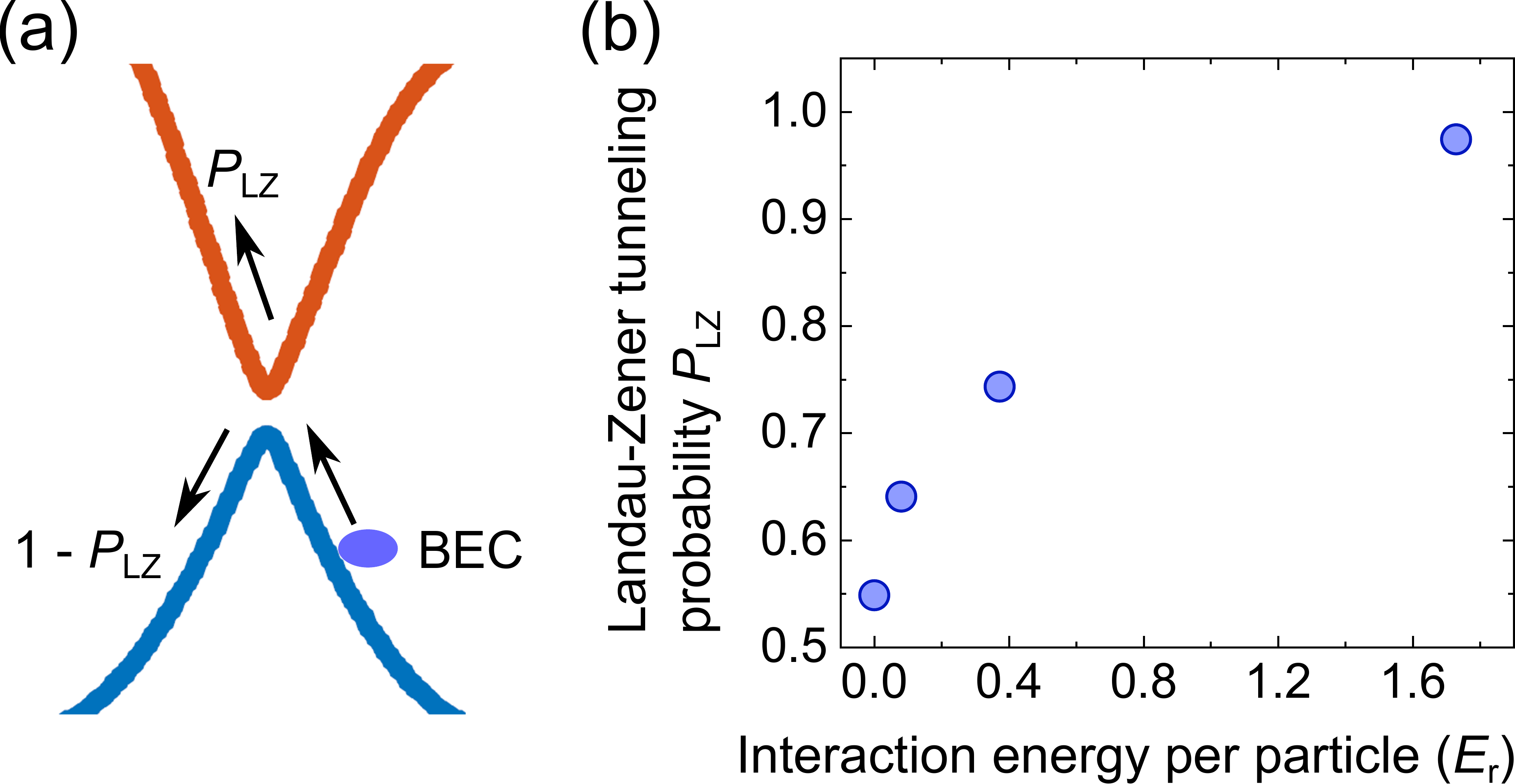}
\caption{\small{
Simulated effects of inter-particle interactions on Landau-Zener tunneling.
(a) Schematic showing atoms in the ground band tunnel to the excited band with a Landau-Zener tunneling probability of $P_\text{LZ}$.
(b) Calculated $P_\text{LZ}$ increases and eventually approaches 1 with increasing interaction energy per particle. 
In the simulation, 15000 $^{87}$Rb atoms are prepared in a shallow 1D harmonic trap, where the trap frequency is 25 Hz. The single-particle gap size is 0.14 $\E_r$. Here $d q_y/d t = 0.188$ K/ms.}
}
\label{FigSI_LZ}
\end{figure}

While in our experiment it is difficult to control the inter-particle interactions
of $^{87}$Rb by adjusting its scattering length, 
we have performed GP simulations to explore how (repulsive) inter-particle interactions may affect atoms' quantum transport in a gapped band when RF couplings are applied. 
In the simulation, 15000 $^{87}$Rb atoms at zero temperature are prepared in a shallow 1D harmonic trap (along the Raman beam or transport direction $\hat{y}$), where the trap frequency is 25 Hz. 
Such a 1D trap is not available in our current experimental setup but is chosen for our simulation setup to enhance and demonstrate the interaction effects that could be explored in future experiments,
and for the relative ease of simulation (where we can use the 1D GP equation rather than the more time-consuming 3D GP equation but still demonstrate the essential physics).
In principle, the 1D trap can be experimentally realized by applying a 2D optical lattice  that realizes parallel 1D tubes, where inter-particle interactions can be tuned by adjusting the lattice confinement.
To reduce computation time and without loss of generality, this simulation adopts the 1D GP equation in a four-spin model (see below for details of four-spin and eight-spin models) with otherwise similar parameters used for the simulation in Fig.~\ref{Fig3} except that the RF couplings are applied such that the single-particle band gap is 0.14 $\E_r$.
In the simulation the interaction energy per particle is varied by varying an effective 1D scattering length (or coupling parameter $g_\text{1D}$), 
which could, for example, be varied by the confinement potential if the 1D tubes are realized by a 2D optical lattice.
Applying a force leads to quantum transport of the atoms (with $d q_y/d t = 0.188$ K/ms), which are initially prepared at the ground band and now tunnel to the excited band with a probability of $P_\text{LZ}$ [Fig.~\ref{FigSI_LZ}(a)].
After the atoms pass the gap, we calculate the fraction of atoms in the excited band, i.e., $P_\text{LZ}$.  
As shown in Fig.~\ref{FigSI_LZ}(b), the calculated $P_\text{LZ}$ increases and eventually approaches 1 with increasing interaction energy per particle.
This means that a strong interaction could change the topology underlying the transport in the momentum space, making atoms effectively experience a M\"obius strip (band crossings) rather than a regular strip (gapped band structure) experienced by a non-interacting BEC.

\section{\label{appendix::H_8spins} Hamiltonians including eight spin states}
\begin{figure}[h]
\includegraphics[width=3.4in]{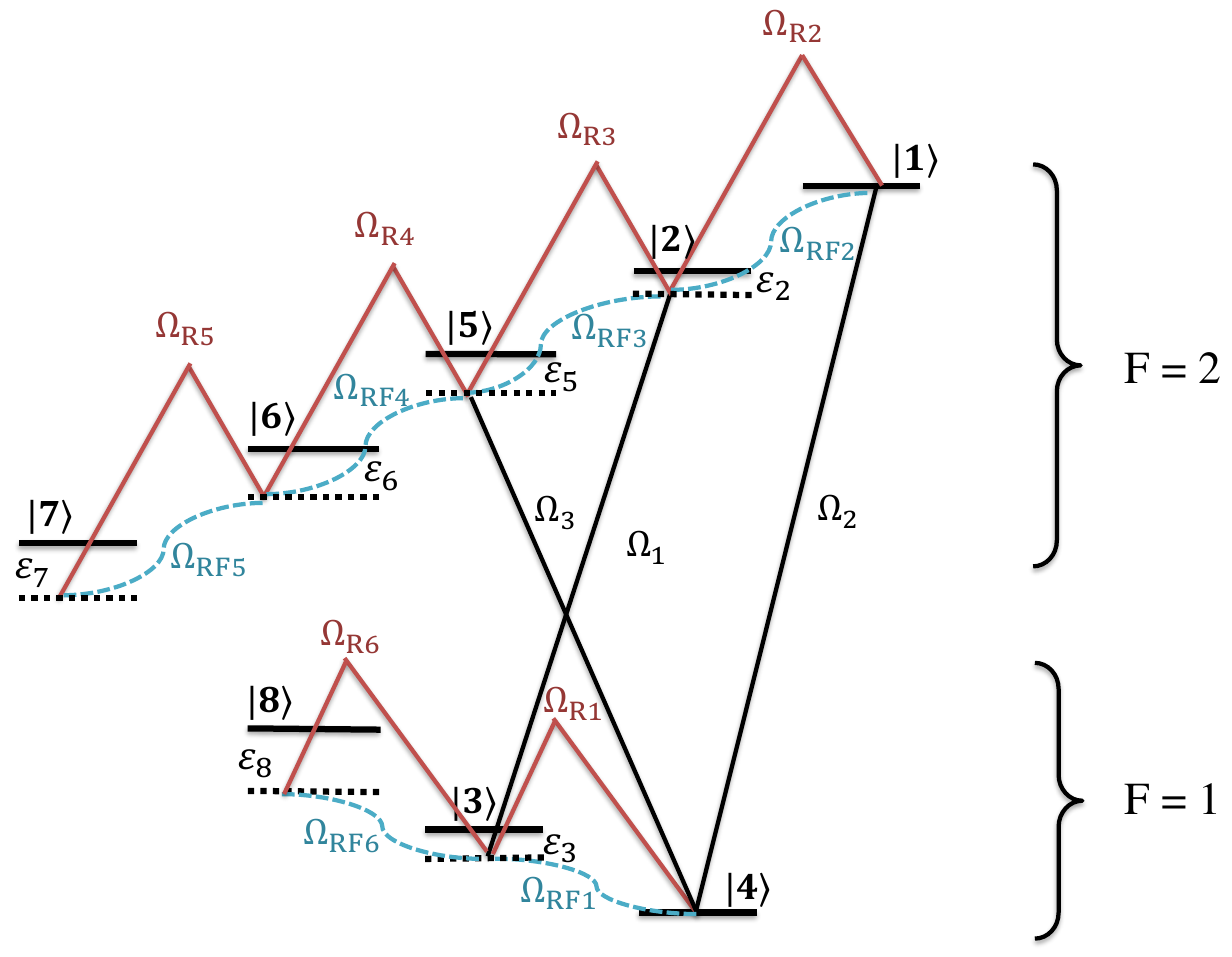}
\caption{\small{
Level diagram of $\mathbf{^{87}}$Rb and light couplings.
Schematic of the eight spin states $\ket{1}$,...,$\ket{8}$ in the $F=1$ and $F=2$ 
hyperfine manifolds. Here $\Omega_\text{R1,...,R6}$ are Raman couplings, $\Omega_{1,2,3}$ are microwave couplings, $\Omega_\text{RF1,...,RF6}$ are RF couplings, and $\varepsilon_{1,..,8}$ are respective ``site energies'' for spin states $\ket{1}$,...,$\ket{8}$ in the synthetic dimension.
}
}
\label{FigSI_8_level}
\end{figure}

\noindent
(1) \textit{The eight-spin model}\\
{When the Zeeman splitting is not big enough, the four states discussed in the main text may not be completely decoupled from other hyperfine spin states. It is thus desirable to investigate effects of those extra spin states. In this section, we consider the effects of the additional ground hyperfine spin states ($\ket{5}=\ket{2,0}$, $\ket{6}=\ket{2,-1}$, $\ket{7}=\ket{2,-2}$, $\ket{8}=\ket{1,-1}$) other than the four spin states $\ket{1}$, $\ket{2}$, $\ket{3}$, $\ket{4}$ we have focused on so far.} 
Figure \ref{FigSI_8_level} shows all the eight spin states in the $F=1$ and $F=2$ 
hyperfine manifolds of $^{87}$Rb, where $\Omega_\text{R1,...,R6}$ are Raman couplings, $\Omega_{1,2,3}$ are microwave couplings, and $\Omega_\text{RF1,...,RF6}$ are RF couplings.
Note that {in our experiment} these Raman couplings are delivered from the same pair of Raman lasers, and these RF couplings come from the same RF wave.
On the other hand, microwave couplings $\Omega_{1,2}$ respectively originate from microwaves 1 and 2 of different frequencies.
Note that microwave 1 can also induce a microwave coupling $\Omega_3$ because the energy splitting between $\ket{2}$ and $\ket{3}$ slightly differs from that between $\ket{4}$ and $\ket{5}$ by $\varepsilon_5$.
However, we empirically find the upper bound of $|\Omega_3/\Omega_1|$ in our setup is only $0.2$, which makes the effects of $\Omega_3$ unimportant, as verified by the detailed calculations below. 

In units of $\E_r$, $\varepsilon_1=0$, $\varepsilon_2=\varepsilon_0=2.4$, $\varepsilon_3=\varepsilon_0=2.4$, $\varepsilon_4=0$, $\varepsilon_5=3.8$, $\varepsilon_6=4.2$, $\varepsilon_7=3.7$, and $\varepsilon_8=5.7$ are determined by the Zeeman splittings and frequencies of the light fields.
They are respective ``site energies'' for spin states $\ket{1}$,...,$\ket{8}$ in the synthetic dimension.

Previously, we only considered spin states $\ket{1}$, $\ket{2}$, $\ket{3}$, and $\ket{4}$ in Hamiltonians.
This {adequately} describes the physics presented in this work.
However, since $\varepsilon_{5,6,7,8}$ are not large enough, spin states $\ket{5}$, $\ket{6}$, $\ket{7}$, and $\ket{8}$ cannot be completely neglected.
In fact, we have included such extra spins into both the eigenstate and GP calculations to better capture the transport experiments quantitatively, as explained below.

Consider the Hamiltonian $H'$ in Eq.~(\ref{H'}) with zero detunings. 
Based on Fig.~\ref{FigSI_8_level}, we can extend the 4-by-4 matrix in $H'$ to an 8-by-8 matrix $H'_{8\times8}$ written in the basis of $\{\ket{1},...,\ket{8}\}$.
The matrix elements $\langle i|H'_{8\times8}|j\rangle$, $i,j=1,2,3,4$, are exactly identical to those in $H'$. 
The extra nonzero matrix elements are:\\  
$\langle 2|H'_{8\times8}|5\rangle=\langle 5|H'_{8\times8}|2\rangle^*
=\frac{\Omega_\text{R3}}{2}e^{-iKy}e^{i\theta_\text{R}}
+\frac{\Omega_\text{RF3}}{2}e^{-i\theta_\text{RF}}$,\\
$\langle 3|H'_{8\times8}|8\rangle=\langle 8|H'_{8\times8}|3\rangle^*
=\frac{\Omega_\text{R6}}{2}e^{iKy}e^{-i\theta_\text{R}}
+\frac{\Omega_\text{RF6}}{2}e^{i\theta_\text{RF}}$,\\
$\langle 4|H'_{8\times8}|5\rangle=\langle 5|H'_{8\times8}|4\rangle^*
=\frac{\Omega_\text{3}}{2}e^{i\theta_\text{3}}$,\\
$\langle 5|H'_{8\times8}|5\rangle=\varepsilon_5$,\\
$\langle 5|H'_{8\times8}|6\rangle=\langle 6|H'_{8\times8}|5\rangle^*
=\frac{\Omega_\text{R4}}{2}e^{-iKy}e^{i\theta_\text{R}}
+\frac{\Omega_\text{RF4}}{2}e^{-i\theta_\text{RF}}$,\\
$\langle 6|H'_{8\times8}|6\rangle=\varepsilon_6$,\\
$\langle 6|H'_{8\times8}|7\rangle=\langle 7|H'_{8\times8}|6\rangle^*
=\frac{\Omega_\text{R5}}{2}e^{-iKy}e^{i\theta_\text{R}}
+\frac{\Omega_\text{RF5}}{2}e^{-i\theta_\text{RF}}$,\\
$\langle 7|H'_{8\times8}|7\rangle=\varepsilon_7$,\\
$\langle 8|H'_{8\times8}|8\rangle=\varepsilon_8$.

Theoretically, for the Raman couplings, $\Omega_\text{R6}/\Omega_\text{R1}=1$,
$\Omega_\text{R2,R5}/\Omega_\text{R1}=-\sqrt{2}$,
$\Omega_\text{R3,R4}/\Omega_\text{R1}=-\sqrt{3}$.
Such scaling relations also hold for the RF couplings.
For the microwave couplings, $\Omega_2/\Omega_1=\sqrt{2}$.
We have checked that the above relations are consistent with our experimental measurements.
For transport experiments, we typically measure $\Omega_\text{R1}$, $\Omega_\text{R2}$, $\Omega_1$, $\Omega_2$, $\Omega_\text{RF1}$ and use these values and the above scaling relations to obtain other couplings for calculations.

Similarly, we can extend all the previous four-spin Hamiltonians to the corresponding eight-spin Hamiltonians. 
For the parameter regime in this work, we find that including extra spins can
notably modify the shape of band structures and thus atoms' transport. This is the main reason why we need to use eight-spin Hamiltonians for both the eigenstate and GP calculations. 
On the other hand, in both experiments and calculations, the fractional occupation of extra spins is typically small, with an estimated maximum of 5\%.
Therefore, for all the calculated results shown in figures, 
we calculate the fractional population of spin $i$ ($i=1,2,3,4$) as $n_i=N_i/(\sum_{i=1,..,4}N_i)$ 
and the average momentum as $\sum_{i=1,..,4}n_i p_i$, 
where $N_i$ and $p_i$ are respectively the population and mechanical momentum of spin $i$.

In summary, the presence of extra spins mainly modifies the band structures and atoms' transport quantitatively. 
On the other hand, the occupation of these extra {spin states} is typically small. 
{The four-spin model is a good approximation and adequately captures the key physics we study here. }

\noindent
(2) \textit{Symmetry in the eight-spin model}\\
Similar to the four-spin model as discussed above, 
the generalized inversion symmetry is also broken in the eight-spin model.
On the other hand, following similar arguments as for the four-spin model, we show that the eight-spin model still has the nonsymmorphic symmetry when RF couplings are zero.
Recall the nonsymmorphic symmetry operator $\hat{S}=\hat{U}\hat{G}$.
In the eight-spin model, 
the unitary transformation operator $\hat{U}$ becomes
$\ket{1}\bra{1}-\ket{2}\bra{2}-\ket{3}\bra{3}+\ket{4}\bra{4}+\ket{5}\bra{5}-\ket{6}\bra{6}+\ket{7}\bra{7}+\ket{8}\bra{8}$.
Acting $\hat{G}$ ($\hat{G}^{-1}$) to the left (right) of $H_{8\times8}$ ($H'_{8\times8}$ with zero RF couplings) flips the sign of the second, third, sixth rows (columns).
Acting $\hat{U}$ ($\hat{U}^{-1}$) to the left (right) of
$\hat{G}H\hat{G}^{-1}$ flips the sign of the second, third, sixth rows (columns) again.
Thus, $\hat{U}\hat{G}H_{8\times8}\hat{G}^{-1}\hat{U}^{-1}=H_{8\times8}$, 
i.e., $H_{8\times8}$ is invariant under the nonsymmorphic symmetry.
The nonsymmorphic symmetry guaranties band crossings.
However, like the four-spin model, the broken generalized inversion symmetry makes band crossings occur not at the edge of the first Brillouin zone.



\providecommand{\noopsort}[1]{}\providecommand{\singleletter}[1]{#1}%

\end{document}